\documentclass[apjl]{emulateapj}
\usepackage{comment}
\usepackage{ifthen}
\usepackage{amsmath}
\usepackage{amsfonts}
\usepackage{amssymb}

\newboolean{emulateapj}
\setboolean{emulateapj}{true}

\newboolean{astroph}
\setboolean{astroph}{true}

\shortauthors{Kipping \& Bakos}
\shorttitle{Analysis of TrES-2b}
\ifthenelse{\boolean{emulateapj}}{
    \newcommand{\titledag}{$\dagger$}
}{
    \newcommand{\titledag}{\dagger}
}

\begin{document}

\title{Analysis of Kepler's short-cadence photometry for TrES-2b\altaffilmark{\titledag}}

\author{
	{\bf David Kipping$^{1,2}$ \& G\'asp\'ar Bakos$^1$}
}
\altaffiltext{1}{Harvard-Smithsonian Center for Astrophysics,
	Cambridge, MA, dkipping@cfa.harvard.edu}

\altaffiltext{2}{University College London, Dept. of Physics,
 Gower St., London, WC1E 6BT}

\altaffiltext{$\dagger$}{
Based on archival data of the \emph{Kepler} telescope. 
}


\begin{abstract}

We present an analysis of 18 short-cadence (SC) transit lightcurves of TrES-2b 
using quarter 0 (Q0) and quarter 1 (Q1) from the \emph{Kepler Mission}. The 
photometry is of unprecedented precision, 237\,ppm per minute, allowing for the 
most accurate determination of the transit parameters yet obtained for this 
system. Global fits of the transit photometry, radial velocities and known 
transit times are used to obtain a self-consistent set of refined parameters for 
this system, including updated stellar and planetary parameters. Special 
attention is paid to fitting for limb darkening and eccentricity. We place
an upper limit on the occultation depth to be $<72.9$\,ppm to 3-$\sigma$
confidence, indicating TrES-2b has the lowest determined geometric albedo for
an exoplanet, of $A_g<0.146$.

We also produce a transit timing analysis using \emph{Kepler}'s 
short-cadence data and demonstrate exceptional timing precision at the level of 
a few seconds for each transit event. With 18 fully-sampled transits at such 
high precision, we are able to produce stringent constraints on the presence of 
perturbing planets, Trojans and extrasolar moons. We introduce the novel use
of control data to identify phasing effects. We also exclude the previously 
proposed hypotheses of short-period TTV and additional transits but find the
hypothesis of long-term inclination change is neither supported nor refuted by 
our analysis.

\end{abstract}

\keywords{
	planetary systems ---
	stars: individual (TrES-2b) 
	techniques: spectroscopic, photometric
}


\section{INTRODUCTION}
\label{sec:intro}

TrES-2b is a transiting planet discovered by the Trans-atlantic Exoplanet Survey 
(TrES) which happens to reside in the field-of-view for the 
\emph{Kepler Mission}\footnote{http://www.kepler.nasa.gov/sci} 
\citep{basri:2005,koch:2007}. The fact that the planet was discovered 
by TrES \citep{donovan:2007} provides several advantages for \emph{Kepler}. 
Firstly, the star was selected as one of the 512 targets for immediate 
short-cadence (SC) observations. Secondly, the planet's ephemeris is 
well-characterized from ground-based measurements meaning a search for long-term 
transit time variations (TTV) is possible. Thirdly, TrES targeted brighter stars 
than \emph{Kepler} and thus TrES-2 is somewhat brighter (V=11.4) than typical 
\emph{Kepler} stars (V = 12 to 14).

In \citet{gilliland:2010}, a presentation of the first TrES-2b lightcurves was 
provided, but the focus of the paper was to demonstrate the properties of the 
SC data rather than a detailed study of the planet's properties. In this paper, 
we present a comprehensive analysis of the first 18 transits observed by 
\emph{Kepler} in quarter 0 (Q0) and quarter 1 (Q1) in short-cadence mode. The 
photometry is analyzed in combination with the radial velocity (RV) data and 
known transit times of the system, and the combined results allow for a refined 
YY-isochrone analysis \citep{yi:2001} to derive a complete and self-consistent 
set of system parameters. Particular attention is paid to fitting for both 
eccentricity and limb darkening coefficients, making the results as 
model-independent as possible.

TrES-2 is somewhat remarkable, if nothing else, for having been the subject of 
numerous tentative detections. For example, \citet{raetz:2009} claimed to have 
detected repeated dips in the lightcurve 1-2\,hours after the main transit event 
and proposed a second resonant planet as an explanation. \citet{rabus:2009} 
claimed to have detected transit timing variations for TrES-2b of period 
0.21\,cycles (where one cycle is one orbital period of TrES-2b) and 50\,s 
amplitude and proposed a 52\,$M_{\oplus}$ exomoon as a possible explanation. 
Finally, \citet{mislis:2009} claim to have detected long-term inclination change 
in the system. In this work, we will also investigate the compatibility of these 
claims with the \emph{Kepler} photometry.

The SC mode was made available for the purposes of studying asteroseismology and 
transit timing variations (TTV). Although \citet{kipbak:2010} have shown that 
even the long-cadence (LC) is capable of performing TTV at the level of 
$\sim20$\,s, the SC data has the potential for further improvement on this. 
Such precision would allow for the detection of satellites 
\citep{sartoretti:1999,kipping:2009}, Mars-mass perturbing planets 
\citep{agol:2005,holman:2005} and Trojan bodies \citep{ford:2007}.

\section{DATA HANDLING}
\label{sec:data}

In this section, we will list the sequential steps we took in processing the 
\emph{Kepler} photometry for TrES-2b.

\subsection{Data Acquisition}
\label{sub:dataacquisition}

We make use of the ``Data Release 5'' (DR5) from the \emph{Kepler Mission}, 
which consists of quarter 0 (Q0) and quarter 1 (Q1). Full details on the data 
processing pipeline can be found in the DR5 handbook. Numerous improvements have 
been made over the previously available MAST (Multimission Archive at STScI) 
data, including most relevant for this study an inclusion of BJD (Barycentric 
Julian Date) time stamps for each flux measurement. The previous version of the 
data only included cadence numbers and thus the inclusion of barycentric 
corrected time stamps is a marked improvement\footnote{We thank Ron Gilliland 
for useful advise on this topic}.

\subsection{Long-Term Detrending with a Cosine Filter}
\label{sub:detrending}

We make use of the ``raw'' (labelled as ``AP\_RAW\_FLUX'' in the header) data 
processed by the DR5 pipeline and a detailed description can be found in the 
accompanying release notes. The ``raw'' data has been processed using PA 
(Photometric Analysis), which includes cleaning of cosmic ray hits, 
Argabrightenings, removal of background flux, aperture photometry and 
computation of centroid positions.

The data release also includes corrected fluxes (labelled as ``AP\_CORR\_FLUX'' 
in the header), which are outputted from the PDC (Pre-search Data Conditioning) 
algorithm developed by the DAWG (Data Analysis Working Group). As detailed in 
DR5, this data is not recommended for scientific use, owing to, in part, the 
potential for under/over-fitting of the systematic effects.

For the sake of brevity, we do not reproduce the details of the PA and PDC
steps here, but direct those interested to \citet{gilliland:2010} and the DR5 
handbook.

\begin{figure*}
\begin{center}
\includegraphics[width=16.8 cm]{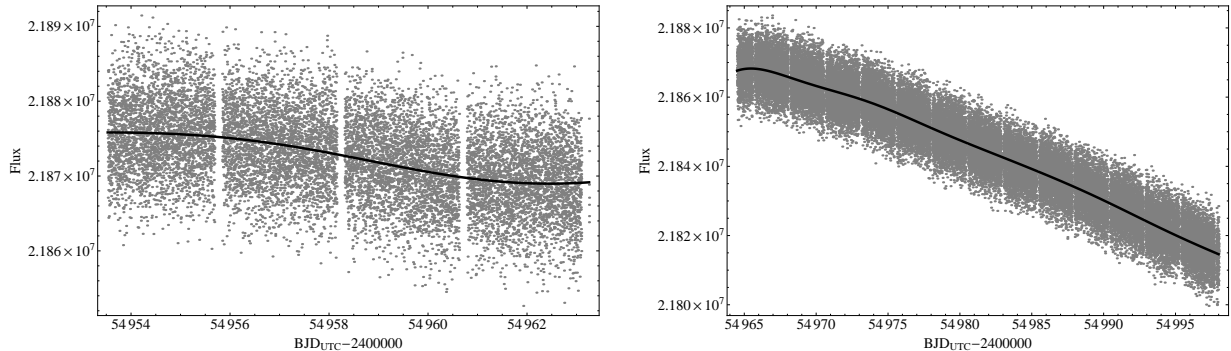}
\caption{\emph{``Raw'' (PA output) flux from DR5 of the \emph{Kepler} pipeline 
for Q0 (left panel) and Q1 (right panel) of the star TrES-2. Overlaid is our 
model for the long-term trend, computed using a discrete cosine transform for 
each data set. Outliers have been removed.}} 
\label{fig:Q0Q1raw}
\end{center}
\end{figure*}

The Q0 and Q1 PA photometry are shown in Figure~\ref{fig:Q0Q1raw}
respectively. One challenge in attempting a correction is assessing which 
components are astrophysical in nature and which are instrumental. In general, 
we wish to preserve the astrophysical signal as much as possible. However, in 
practice, any signals occurring on timescales greater than the orbital period of 
the transiting planet, whether instrumental or astrophysical, have a negligible 
impact on the morphology of the transit lightcurve, which is ultimately what we 
are interested in for this study. These signals may be removed by applying a 
high-pass filter to the photometry, in a similar way as was used by 
\citet{mazeh:2010} for the spaced-based CoRoT photometry.

To remove the long-term trend, visible in 
Figure~\ref{fig:Q0Q1raw}, we applied a discrete cosine transform 
\citep{ahmed:1974} adopted to the unevenly spaced data.

We first removed the 18 transit events with a margin of 6559.4\,s either side of 
the times of transit minimum. This value was chosen as it represents the 
1st-to-4th contact duration and thus we essentially remove the transit plus
one half of the total transit duration either side of each event. We also remove
outliers, identified as those points lying 3-$\sigma$ away from a
spline-interpolated running median of window-size 20\,minutes. Treating
Q0 and Q1 separately, we fitted the remaining data with a linear combination of 
the first N low-frequency cosine functions:

\begin{align}
f_i(t_j) = \cos\Big(\frac{2\pi t_j i}{2D}\Big)
\label{eqn:cosinefilter}
\end{align}

Where $t_j$ is the timing of the $j^{\mathrm{th}}$ measurement, $i=0,N$ in 
integer steps and $N$ is equal to the rounded integer value
of $(2D/4P_P)$ where $D$ is the timespan of the observations and $P_P$ is the
orbital period of TrES-2b. For Q0, we used $N=2$ and for Q1, $N=7$. We then
fit for the linear coefficient, $a_i$, for each of the cosine functions, so that
the fitted model is:

\begin{align}
\mathcal{M}(t_j) = \sum_{i=0}^N a_i f_i(t_j)
\label{eqn:trendmodel}
\end{align}

We then subtracted model $\mathcal{M}$ from the lightcurve (including the 
transits). The model is shown over the data for the Q0 and Q1 photometry in 
Figure~\ref{fig:Q0Q1raw}.

\subsection{Median Normalizations}
\label{sub:normalizations}

A second stage of normalization is applied to the data after the long-term
detrending. Here, we split the lightcurve up into 18 individual transit and
occultation events (giving 36 arrays in total). Each array spans from -0.125 to
+0.125 in orbital phase surrounding the event in question. The fluxes and
associated errors in each array and then divided by the median of each array. 
This is similar to the technique adopted by \citet{kipbak:2010}.

\subsection{Outliers}
\label{sub:outliers}

Despite the PA processing, some outliers still remain in our detrended,
normalized photometry. We must remove these before it is possible to perform the 
final lightcurve fits. Since these outliers can occur within the transit event 
itself, it is necessary to perform a preliminary fit of the transits and then 
remove outliers from the residuals.

For the purpose of the identifying outliers, we perform an initial global fit
(as described later in \S~\ref{sub:fitting}). The residuals are
then used to search for outlier points by flagging those which occur 3-$\sigma$
away from the model.

\subsection{Time Stamps}
\label{sub:timestamps}

In the DR5 handbook, the following advise is given:

``The advice of the DAWG [Data Analysis Working Group] is not to consider as 
scientifically significant relative timing variations less than the read time 
(0.5s) or absolute timing accuracy better than one frame time (6.5s) until such 
time as the stability and accuracy of time stamps can be documented to near the 
theoretical limit.''

Relative time differences correspond to, for example, performing TTV (transit
timing variations) and TDV (transit duration variations) on 
the \emph{Kepler} data alone. Absolute time differences corresponds to, for 
example, performing TTV and TDV on the \emph{Kepler} data plus all previously 
observed data. We stress these limitations early on in our study. 

The \emph{Kepler} time stamps from DR5 are in BJD$_{\mathrm{UTC}}$ (Barycentric
Julian Date in Coordinated Universal Time) and a correction to 
$\mathrm{BJD}_{\mathrm{TDB}}$ (Barycentric Julian Date in Terrestrial Dynamic 
Time) is advocated by \citet{eastman:2010} in all transit timing studies. The 
correction between UTC and TDB is given by 
$\mathrm{BJD}_{\mathrm{TDB}} = \mathrm{BJD}_{\mathrm{UTC}} + N + 32.184$, where 
$N$ is the number of leap seconds which have elapsed since 1961. This correction
has been applied all data analyzed in this study.

\subsection{Correlated Noise}
\label{sub:rednoise}

Time-correlated noise may affect the estimation of lightcurve parameters 
\citep{carter:2009b} and so we here discuss the degree to which this data set is 
affected by correlated noise. We present two methods of assessing the degree
of red noise, following the approach adopted by \citet{carter:2009a}.

\begin{figure*}
\begin{center}
\includegraphics[width=16.8 cm]{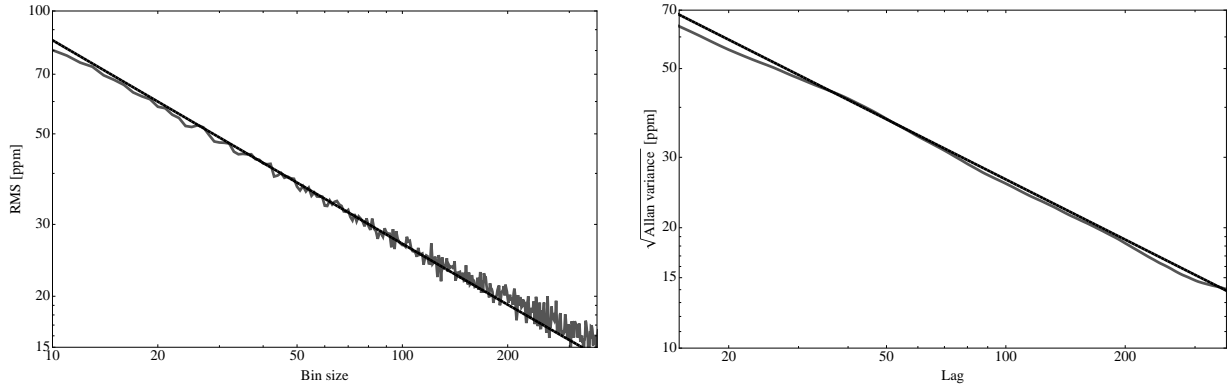}
\caption{\emph{Assessment of correlated noise. Left panel: The r.m.s. of the
time-binned residuals as function of bin-size. Right panel: The square-root
of the Allan variance of the residuals, as a function of lag. The gray lines
are computed from the data and the black lines show the expected behavior for
uncorrelated noise.}}
\label{fig:correl}
\end{center}
\end{figure*}

First, using the residuals of our final fits (which will be introduced in more 
detail in \S\ref{sub:fitting}), we bin the residuals into a bin size $N$ and 
evaluate the r.m.s. of the data. We repeat this process from $N=1$ up to 
$N=360$ (which is approximately equal to the time span of each discrete 
lightcurve array, 0.25\,days) and the results are shown in 
Figure~\ref{fig:correl}. The figure reveals that our corrected data is follows 
closely the expectation of independent random numbers, 
$\sigma_N = \sigma_1 N^{-1/2} [M/(M-1)]^{1/2}$, where $M$ is the number of bins.

Second, we computed the \citet{allan:1966} variance $\sigma_A^2(l)$ of the
residuals, defined as:

\begin{align}
\sigma_A^2(l) = \frac{1}{2(N+1-2l)} \sum_{i=0}^{N-2l} \Bigg( \frac{1}{l} \sum_{j=0}^{l-1} r_{i+j}-r_{i+j+l}\Bigg)^2
\label{eqn:allan}
\end{align}

where $r_k$ denotes the residual of the $k^{\mathrm{th}}$ data points, $N$ is
the number of data points and $l$ is the lag. For independent residuals, one
expects $\sigma_A^2(l)\simeq\sigma_A^2(0)/l$, for which our residuals can be
seen to be satisfy in Figure~\ref{fig:correl}.

The bulk r.m.s. of our data set is 237.2\,ppm. In comparison, the PCD corrected
photometry has a bulk r.m.s. of 230.5\,ppm (after removing outliers). It is 
possible that PCD overfitted the data or that our own correction is an underfit. 
Based upon the analysis above though, we find no strong evidence for correlated 
noise in our corrected photometry, which would be expected for underfitted 
detrending.

\section{MODEL DETAILS}
\label{sec:model}

\subsection{Model Generation}
\label{sub:modelgen}

\subsubsection{Lightcurves}
\label{subsub:lightcurves}

The primary transit lightcurve model is computed using the \citet{mandel:2002} 
limb darkening algorithm. The outputted fluxes are corrected for variable 
baseline flux, OOT, to propagate the baseline r.m.s. into the lightcurve. 
The occultation lightcurve is computed in the same way, except the limb 
darkening coefficients are fixed to zero and the final lightcurve is then 
squashed by a factor which is equal to the ratio of the transit to occultation 
depth. We note that multiplying the ratio-of-radii squared, $p^2$, by this 
factor and then feeding this value into the \citet{mandel:2002} code instead 
would be erroneous, since the algorithm would think the planet was very small 
leading to sharper ingress/egress features. By applying the transformation at 
the end, we preserve the correct lightcurve morphology.

The true anomaly is calculated from the time stamps by solving Kepler's Equation
at every instance. Transit durations are computed using the 
expressions of \citet{kipping:2008}, which account for orbital eccentricity. 
Although TrES-2b is believed to be on a circular orbit \citep{donovan:2007}, 
using the most general equations allows us to float the eccentricity parameters 
to propagate their uncertainties. Recent \emph{Spitzer} occultation
measurements by \citet{donovan:2010} strongly constrain 
$e\cos\omega = (0.00053 \pm 0.00102)$. We allow both $e\sin\omega$ and 
$e\cos\omega$ to be fitted for in our global fits, but as the $e\cos\omega$ term 
moves away from the value found by \citet{donovan:2010}, a $\chi^2$ penalty is 
assigned (see Equation~\ref{eqn:merit})

We initially use the errors from the normalized PA lightcurve and then
rescale the errors after one iteration of the global fit. Errors are scaled
such that the $\chi^2$ function for the transit data, the occultation data,
the RV data and the transit times are each equal to the number of data points
in that fit minus the number of degrees of freedom in the model.

\subsubsection{Radial velocity}
\label{subsub:rv}

The radial velocity curve is computed assuming a single planet in a Keplerian 
orbit. The free parameters in the model are the time of transit, the orbital 
period, $e\cos\omega$, $e\sin\omega$ and the semi-amplitude $K$. We do not 
consider the Rossiter-McLaughlin (RM) effect since our principal goal is to 
characterize the orbit and 
the points for the RM lead to very little improvement in the parameters listed 
here \citep{winn:2008}, but severe increases in CPU time. As a result, we only 
use the radial velocities from \citet{donovan:2007}, taking care to convert the 
times to BJD$_{\mathrm{TDB}}$.

\subsubsection{Transit times}
\label{subsub:transittimes}

So far we have three data sets which are fitted for; the transits, the 
occultations and the radial velocities. Usefully, TrES-2b is a relatively 
old discovery and several years of transit measurements exist. However, most of 
these come from amateur measurements, which have not been peer-reviewed, and 
thus may not be as reliable. A resolution to this is to use median statistics to 
define the merit function and therefore provide a robust estimation of the 
goodness of fit, even in the presence of outliers. We therefore add in a fourth 
data set to our global fits coming from the timings, which provides extremely 
tight constraints on the ephemeris.

Let us consider the typical merit of function first, which is based on mean 
statistics. In order to compute the ephemeris, we throw in a trial model of 
$\tau + n P$ and then calculate the residuals for each point, $r_i$. 
We then evaluate the weighted squares of each of these measurements, given by 
$(r_i/\sigma_i)^2$, where $\sigma_i$ is the measurement error. In a normal 
analysis, we would then sum these weighted squares together to give the $\chi^2$ 
and then perturb the model until we obtain the lowest possible $\chi^2$:

\begin{align}
\chi^2 &= \sum_{i=1}^{n} (r_i/\sigma_i)^2 \nonumber \\
\chi^2 &= n \, \frac{\sum_{i=1}^{n} (r_i/\sigma_i)^2}{n} \nonumber \\
\chi^2 &= n \, \mathrm{Mean}\{(r_i/\sigma_i)^2\}
\end{align}

Inspection of the above equation reveals the simple way in which we can change 
our merit function to be in the form of median statistics, to give ``$\xi^2$'':

\begin{equation}
\xi^2 = n \, \mathrm{Median}\{(r_i/\sigma_i)^2\}
\end{equation}

The $\xi^2$ distribution is very similar to that of the $\chi^2$ distribution, 
but a scaling factor is required to make them equivalent. This factor is 
frequently required when converting median statistics to mean statistics; for 
example the standard deviation is given by 1.4286 times by the 
median-absolute-deviation and the error on the sampling median is 1.253 times 
the error on the sampling mean. In this case, the factor was computed using 
Monte Carlo simulations where we found the factor $2.2\dot{6}$ provides the 
correct scaling.

Times of transit minimum found in the exoplanet literature and the ETD 
(Exoplanet Transit Database\footnote{http://var2.astro.cz/ETD/}) are almost 
always in HJD$_{\mathrm{UTC}}$ (Heliocentric Julian Date in Coordinated 
Universal Time). We use the JPL Horizons ephemeris to convert the 
HJD$_{\mathrm{UTC}}$ times to BJD$_{\mathrm{UTC}}$ and then apply the correction 
for leap-seconds to yield BJD$_{\mathrm{TDB}}$. The list of used transit times 
is presented in the appendix, Table~\ref{tab:allttv}.

\subsection{Fitting Algorithm}
\label{sub:fitting}

Fits are accomplished by using a Metropolis-Hastings Markov Chain Monte Carlo 
(MCMC) algorithm \citep{tegmark:2004,holman:2006}. The
routine begins from a starting point, which we select to be 5-$\sigma$ away from
the estimated solution, and then generates new trial parameters by making a jump
computed using a Gaussian proposal distribution centered upon 
the current position with a standard deviation given by the ``jump size''. Jump
sizes are selected, usually through a process of iteration, to be equal to the
1-$\sigma$ uncertainties for each parameter.

The trial parameters are then used to produce a model, which is compared to the
observations to produce the goodness-of-fit merit function, $\chi^2$. Trials
producing a lower $\chi^2$ than the current position are always accepted and the
trial position becomes the current position, constituting an accepted jump.
Trials producing a higher $\chi^2$ are accepted with a probability:

\begin{align}
\mathrm{P}(\mathrm{accept}) &= \exp(-\Delta \chi^2/2)
\label{eqn:metropolis}
\end{align}

where $\Delta \chi^2$ is the difference in $\chi^2$ between the current
position and the trial position. The algorithm stops when 125,000 trials have 
been accepted and the first 25,000 (20\%) are discarded as burn-in leaving 
$10^5$ points for the posterior distributions. Our algorithm follows the same
procedure detailed in \citet{ford:2005}. The overall merit function 
(see \S\ref{sub:modelgen} for details) is given by:

\begin{align}\label{eqn:merit}
\chi^2 =& \sum_{i=1}^{n_P} \Big(\frac{ f_{\mathrm{obs},i}^P - f_{\mathrm{model},i}^P }{\delta f_i^P}\Big)^2 + \sum_{i=1}^{n_S} \Big(\frac{ f_{\mathrm{obs},i}^S - f_{\mathrm{model},i}^S }{\delta f_i^S}\Big)^2 \nonumber \\
\qquad& + \sum_{i=1}^{n_R} \Big(\frac{ v_{\mathrm{obs},i} - v_{\mathrm{model},i} }{\delta v_i}\Big)^2 + \Big( \frac{e\cos\omega - 0.00053}{0.00102} \Big)^2  \nonumber \\
\qquad& + 2.2\dot{6} n_T \mathrm{Median} \Big\{ \Big(\frac{ t_{\mathrm{obs},i} - t_{\mathrm{model},i} }{\delta t_i}\Big)^2 \Big\}
\end{align}

We fit using 14 free parameters \{$\tau$, $P$, $p^2$, $\tilde{T}_1$,
$b$, $e\cos\omega$, $e\sin\omega$, OOT, OOS, $w_1$, $w_2$, $F_P/F_*$, $K$, 
$\gamma$\}, which we elaborate on here. $\tau$ is the time of transit minimum 
for the optimum epoch (that epoch which produces the minimum correlation to $P$) 
and is defined as the instance when the planet-star sky-projected separation is 
minimized (note, this is frequently given the misnomer ``mid-transit time''). 
$P$ is the orbital period, $p^2$ is the ratio-of-radii squared and $b$ is the 
impact parameter (defined as the planet-star sky-projected separation in units
of the stellar radius at the instance of inferior conjunction). $\tilde{T}$ is 
the transit duration between the instance of the planet's center crossing the
stellar limb to exiting under the same condition. $\tilde{T}_1$ is the
``one-term'' approximate expression for this parameter, given by Equation 15 in
\citet{investigations:2010} (an exact analytic form for $\tilde{T}$ is not 
possible, see \citet{investigations:2010} for details). Stellar limb darkening 
is accounted for using a quadratic limb darkening model, modeling the specific 
intensity as a function of $\mu$:

\begin{align}
\frac{I_{\mu}}{I_1} &= 1-u_1(1-\mu) - u_2 (1-\mu)^2
\label{eqn:LD}
\end{align}

where $\mu$ is the cosine of the angle between the observer and the normal
to the stellar surface. $u_1$ and $u_2$ are known to be highly correlated
in typical lightcurve fits \citep{pal:2008} and instead we opt to use $w_1$
and $w_2$, which are related to the quadratic limb darkening coefficients 
\citep{pal:2008} via:

\begin{align}
w_1 &= u_1 \cos\varphi - u_2 \sin\varphi \nonumber \\
w_2 &= u_2 \cos\varphi+ u_1 \sin\varphi
\label{eqn:w1w2}
\end{align}

\citet{pal:2008} has advocated using this linear combination instead of $u_1$
and $u_2$ due to the improved decorrelations. The author also recommends using
$\varphi=40^{\circ}$, which was done so in this study. During the MCMC,
we discard any trials which yield unphysical limb darkening coefficients,
defined as those which are not everywhere positive and produce a monotonically
decreasing profile from limb to center. This is implemented by using the
conditions \citep{carter:2009a}: $u_1+u_2<1$, $u_1+u_2>0$ and $u_1>0$.

Finally, the final parameter values quoted in this paper are given by the
median of all of the accepted MCMC trials for the parameter in question.
Similarly, 1-$\sigma$ uncertainties are calculated by evaluating the 34.1\%
quantiles either side of the median.

\subsubsection{Why fit for eccentricity?}
\label{subsub:whyecc}

Some readers may question why we choose to fit for eccentricity when the orbit 
is consistent with a circular orbit \citep{donovan:2007}. Firstly, we point out 
that by using all of the known transit times, the \emph{Kepler} lightcurves and 
occultation constraints we are able to derive the most precise constraints 
on $e$ yet for this system, which is a worthwhile goal in itself.

However, the most important reason for fitting for $e$ is that any uncertainty 
on $e$ leads to inflated uncertainties on the derived stellar density, $\rho_*$. 
As pointed out by \citet{investigations:2010}, the retrieved stellar density is 
given by the approximation $\rho_* \simeq \rho_{*,\mathrm{circ}}/\Psi(e,\omega)$ 
where the first term is the stellar density derived from a circular fit and 
$\Psi$ is given by:

\begin{equation}
\Psi = \frac{(1+e \sin\omega)^{3}}{(1+e^2)^{3/2}}
\label{eqn:psi}
\end{equation}

$\rho_{*,circ}$ is determined purely photometrically and thus the uncertainty
will decrease as $\sim1/\sqrt{N_{\mathrm{transits}}}$, where 
$N_{\mathrm{transits}}$ is the number of observed transits by \emph{Kepler}.
This parameter can therefore be expected to be known to very high precision
by the end of the \emph{Kepler Mission}, given the short orbital period of
TrES-2b. In contrast, $\Psi$ can only be measured by radial velocities and/or 
occultation events. Given the visible bandpass of \emph{Kepler}, occultation
events are not expected to be detectable for the majority of transiting planets,
and so the radial velocity determination dominates. With typical transiting
planets receiving sparse radial velocity coverage, it can be appreciated
that the uncertainty on $\Psi$ will often be the limiting factor in the
measurement of a precise $\rho_*$.

The point is that we do not know the orbit is exactly circular (indeed this is
practically impossible) and thus we cannot assume $e\sin\omega=0$ and 
$e\cos\omega=0$ exactly. In reality, we have errors on both of these and can 
only say it is circular to within a certain confidence level. This uncertainty 
therefore propagates into a much larger error for the stellar density. As an 
example, \citet{kipbak:2010} compare fits for Kepler-4b through 8b using both 
circular and eccentric fits and find the errors on $\rho_*$ consistently inflate 
for the latter.

\subsubsection{Why fit for limb darkening?}
\label{subsub:whyld}

Another methodology we adopt, which is not a completely standard practice in the 
exoplanet literature, is that we fit for the limb darkening coefficients. 
Fitting for quadratic limb darkening requires a very high signal-to-noise if one 
wishes to achieve convergence, especially for a near-grazing transit. In many 
ground-based measurements, it is not possible to fit for these coefficients, 
although linear limb darkening could be used instead. 

However, if fitting for the limb darkening is viable, it is always preferable. 
This is because transit parameters derived using fixed limb darkening 
coefficients are fundamentally model dependent, where the model is that of the 
stellar atmosphere model. In contrast, transit parameters derived using fitted 
limb darkening are independent of a stellar atmosphere. This makes them vastly 
more robust and reliable.

This point is particularly salient for TrES-2b. For a near-grazing transit, the 
planet only ever crosses the limb, where the star is most severely darkened. 
Thus the choice of limb darkening coefficients has a very significant effect on 
the derived planetary radius and transit depth especially. The total stellar 
flux, which defines the observed transit depth, is essentially extrapolated from 
the stellar centre to the limb based upon the fitted limb darkening coefficients 
of the limb only. This leads to large correlations between the limb darkening 
coefficients and the ratio-of-radii squared.

\subsection{Blending}
\label{sub:blending}

Recently, \citet{daemgen:2009} showed that the TrES-2 has a very nearby star, 
which was proposed to be in binary star system composed of the originally known 
G0 TrES-2A star and a previously undetected K4.5-K6 companion, (labeled TrES2/C 
by the authors). In the z'-band, the magnitude difference was estimated to be 
3.43 and thus we estimate the blending factor $B$ (which is defined in 
\citet{kiptin:2010}) to be $B = (1.04246 \pm 0.00023)$.

This blending acts to dilute the transit depth and thus causes us to 
underestimate the true planetary radius. Correcting for blends may be 
accomplished by following the prescription of \citet{kiptin:2010}, which we 
adhere to in this work. Self-blending due to nightside emission is expected to 
be negligible in the \emph{Kepler} bandpass (see same work) and thus need not be 
accounted for.

\subsection{Limb Darkening Computation}
\label{sub:ldcom}

In \S\ref{sub:ldfit}, we will discuss how limb darkening coefficients are fitted 
for in the final results. However, it is useful to generate the limb darkening 
coefficients from theoretical models for i) providing a sensible starting point 
for the fitting procedure ii) later comparison of theoretical models versus 
fitted limb darkening.

Limb darkening coefficients were calculated for the \emph{Kepler} bandpass for 
TrES-2b. For the \emph{Kepler} bandpass, we used the high resolution 
\emph{Kepler} transmission function found at 
http://keplergo.arc.nasa.gov/CalibrationResponse.shtml. We adopted the 
SME-derived stellar properties reported in \citet{sozzetti:2007}. We employed 
the \citet{kurucz:2006} atmosphere model database providing intensities at 17 
emergent angles, which we interpolated linearly at the adopted 
$T_{\mathrm{eff}}$ and $\log g$ values. The passband-convolved intensities at 
each of the emergent angles were calculated following the procedure in 
\citet{claret:2000}. To compute the coefficients we used the limb darkening
law given in Equation~\ref{eqn:LD}.

The final coefficients resulted from a 
least squares singular value decomposition fit to 11 of the 17 available 
emergent angles. The reason to eliminate 6 of the angles is avoiding excessive 
weight on the stellar limb by using a uniform sampling (10 $\mu$ values from 
0.1 to 1, plus $\mu=0.05$), as suggested by \citet{diaz:1995}.

\subsection{Drifts and Trojans}
\label{sub:trojans}

Before we provide the final results, we discuss how we performed global
fits including a linear drift in the radial velocities, $\dot{\gamma}$, and
a temporal offset between the radial velocity null and the time of transit
minimum, $\Delta t$ (such a temporal offset is expected to be induced by Trojans 
\citep{ford:2006}). By switching on and off these parameters, there
are four possible permutations of the fits we can execute; eight when one
switches on/off eccentricity as well (see Table~\ref{tab:BICs}).

In general, fitting for an excessive number of free parameters is undesirable
as it increases the errors on the other terms. In order to decide whether
these two additional parameters should be included or not, one may evaluate
the Bayesian Information Criterion (BIC) \citep{schwarz:1978,liddle:2007}, for 
each of the proposed models. The model with the lowest BIC is accepted and 
subsequently used in the global fits reported in the next section. These fits 
used 125,000 MCMC trials with a more aggressive $\chi^2$ minimization downhill 
simplex implemented afterwards. It is based on this lowest $\chi^2$ solution 
from which the BIC is computed.

We therefore performed eight versions of our global fits, with the results for
the BIC values presented in Table~\ref{tab:BICs}. We find that neither
a drift nor a temporal offset are accepted for either the circular or eccentric
models. We therefore proceed to consider them fixed to zero. The results,
however, do allow us to place upper limits on $\dot{\gamma}$ and $\Delta t$.
We find $|\dot{\gamma}|<0.12$\,$\mathrm{m}\mathrm{s}^{-1}\mathrm{year}^{-1}$ 
and $|\Delta t|<0.15$\,days to 3-$\sigma$ confidence. This excludes a 
$>0.94$\,$M_J$ Trojan in a 1:1 resonance with TrES-2b.

\begin{table}
\caption{\emph{Bayesian Information Criterion (BIC) values for eight
models executed for the global fits. We highlight the lowest BIC values.}} 
\centering 
\begin{tabular}{c c} 
\hline\hline 
Model & BIC \\ [0.5ex] 
\hline
$e=0$,$|\dot{\gamma}|=0$,$|\Delta t|=0$ & \textbf{34048.7} \\
$e=0$,$|\dot{\gamma}|>0$,$|\Delta t|=0$ & 34064.2 \\
$e=0$,$|\dot{\gamma}|=0$,$|\Delta t|>0$ & 34058.6 \\
$e=0$,$|\dot{\gamma}|>0$,$|\Delta t|>0$ & 34069.7 \\
\hline
$e>0$,$|\dot{\gamma}|=0$,$|\Delta t|=0$ &\textbf{34067.2} \\
$e>0$,$|\dot{\gamma}|>0$,$|\Delta t|=0$ & 34077.5 \\
$e>0$,$|\dot{\gamma}|=0$,$|\Delta t|>0$ & 34259.1 \\
$e>0$,$|\dot{\gamma}|>0$,$|\Delta t|>0$ & 34268.4 \\ [1ex]
\hline\hline 
\end{tabular}
\label{tab:BICs} 
\end{table}

\section{RESULTS OF GLOBAL FITS}
\label{sec:results}

\begin{table*}
\caption{\emph{Results from global fits of TrES-2b using eighteen short-cadence 
(SC) \emph{Kepler} transits. We show results for both circular and eccentric 
fits in columns 2 and 3. In column 4, we provide the previous estimates of the 
system parameters from \citet{donovan:2007}$^{\mathrm{i}}$ 
\citet{holman:2007}$^{\mathrm{ii}}$ and \citet{sozzetti:2007}$^{\mathrm{iii}}$. 
In general, the eccentric fit leads to more realistic errors. Quoted values are 
medians of MCMC trials with errors given by 1-$\sigma$ quantiles. * = fixed 
parameter; $\dagger$ = parameter was floated but not fitted.}} 
\centering 
\begin{tabular}{c c c c} 
\hline\hline 
Parameter & Circular & Eccentric & Previous \\ [0.5ex] 
\hline
\emph{Model indep. params.} & & \\
\hline 
$P$ [days] & $2.47061896_{-0.00000016}^{+0.00000022}$ & $2.47061892_{-0.00000012}^{+0.00000018}$ & $2.470621 \pm 0.000017$ $^{\mathrm{ii}}$ \\
$\tau$ [BJD$_{\mathrm{TDB}}$ - 2,450,000] & $4849.526635_{-0.000026}^{+0.000026}$ & $4849.526640_{-0.000021}^{+0.000022}$ & - \\
$T_{1,4}$ [s] & $6438_{-33}^{+31}$ & $6439_{-28}^{+25}$ & $6624 \pm 72$ $^{\mathrm{ii}}$ \\
$\tilde{T}_1$ [s] & $4624_{-41}^{+42}$ & $4624_{-31}^{+32}$ & - \\
$T_{2,3}$ [s] & $1950_{-110}^{+110}$ & $1942_{-86}^{+84}$ & - \\
$(T_{1,2} \simeq T_{3,4})$ [s] & $2242_{-47}^{+49}$ & $2247_{-37}^{+38}$ & $2459 \pm 162$ $^{\mathrm{ii}}$ \\
$(R_P/R_*)^2$ [\%] & $1.643_{-0.052}^{+0.082}$ & $1.633_{-0.045}^{+0.076}$ & - \\
$b$ & $0.8408_{-0.0053}^{+0.0047}$ & $0.8418_{-0.0045}^{+0.0037}$ & $0.8540 \pm 0.0062$ $^{\mathrm{ii}}$ \\
$\delta_{\mathrm{occultation}}$ [ppm] & $21_{-22}^{+23}$ & $19_{-17}^{+18}$ & - \\
$e\sin \omega$ & $0^{\mathrm{*}}$ & $-0.009_{-0.029}^{+0.024}$ & $0^{\mathrm{*}}$ $^{\mathrm{ii}}$ \\
$e\cos \omega$ & $0^{\mathrm{*}}$ & $0.0005_{-0.0018}^{+0.0018}$ & $0^{\mathrm{*}}$ \\
$\Psi$ & $1^{\mathrm{*}}$ & $0.973_{-0.082}^{+0.071}$ & $1^{\mathrm{*}}$ $^{\mathrm{ii}}$ \\
$K$ [ms$^{-1}$] & $181.4_{-6.7}^{+6.8}$ & $181.0_{-5.4}^{+5.5}$ & $181.3 \pm 2.6$ $^{\mathrm{i}}$ \\
$\gamma$ [ms$^{-1}$] & $-29.9_{-5.6}^{+5.6}$ & $-29.2_{-2.6}^{+2.6}$ & - \\
$B$ & $1.04246 \pm 0.00023$ $\dagger$ & $1.04246 \pm 0.00023$ $\dagger$ & $1^{*}$ $^{\mathrm{ii}}$ \\
$u_1$ & $0.52_{-0.34}^{+0.44}$ & $0.45_{-0.30}^{+0.42}$ & $0.22^{*}$ $^{\mathrm{ii}}$ \\
$u_2$ & $0.06_{-0.48}^{+0.37}$ & $0.12_{-0.46}^{+0.33}$ & $0.32^{*}$ $^{\mathrm{ii}}$ \\
$R_P/R_*$ & $0.1282_{-0.0020}^{+0.0032}$ & $0.1278_{-0.0018}^{+0.0029}$ & $0.1253 \pm 0.0010$ $^{\mathrm{ii}}$ \\
$a/R_*$ & $7.983_{-0.084}^{+0.132}$ & $8.06_{-0.21}^{+0.25}$ & $7.63 \pm 0.12$ $^{\mathrm{ii}}$ \\
$i$ [$^{\circ}$] & $83.952_{-0.094}^{+0.131}$ & $84.07_{-0.31}^{+0.34}$ & $83.57 \pm 0.14$ $^{\mathrm{ii}}$ \\
$e$ & $0^{\mathrm{*}}$ & $0.018_{-0.013}^{+0.023}$ & $0^{\mathrm{*}}$ $^{\mathrm{ii}}$ \\
$\omega$ [$^{\circ}$] & - & $268_{-180}^{+7}$ & - $^{\mathrm{ii}}$ \\
$\rho_*$ [g/,cm$^{-3}$] & $1.63_{-0.13}^{+0.16}$ & $1.63_{-0.13}^{+0.16}$ & $1.375 \pm 0.065$ $^{\mathrm{iii}}$ \\
$\log(g_P\,[\mathrm{cgs}])$ & $3.317_{-0.018}^{+0.018}$ & $3.326_{-0.024}^{+0.025}$ & - \\
\hline
\emph{Model depend. params.} & & \\
\hline
$T_{\mathrm{eff}}$ [K] (SME) & $5850 \pm 50$ $\dagger$ & $5850 \pm 50$ $\dagger$ & $5850 \pm 50$ $^{\mathrm{iii}}$ \\
$\log(g\,[\mathrm{cgs}])$ (SME) & $4.4 \pm 0.1$ $\dagger$ & $4.4 \pm 0.1$ $\dagger$ & $4.4 \pm 0.1$ $^{\mathrm{iii}}$ \\
(Fe/H) [dex] (SME) & $-0.15 \pm 0.10$ $\dagger$ & $-0.15 \pm 0.10$ $\dagger$ & $-0.15 \pm 0.10$ $^{\mathrm{iii}}$ \\
$M_*$ [$M_{\odot}$] & $0.992_{-0.050}^{+0.040}$ & $0.990_{-0.048}^{+0.041}$ & $0.980 \pm 0.062$ $^{\mathrm{iii}}$ \\ 
$R_*$ [$R_{\odot}$] & $0.958_{-0.020}^{+0.018}$ & $0.952_{-0.029}^{+0.028}$ & $1.000_{-0.033}^{+0.036}$ $^{\mathrm{iii}}$ \\
$\log(g\,[\mathrm{cgs}])$ & $4.469_{-0.012}^{+0.015}$ & $4.475_{-0.024}^{+0.024}$ & $4.426_{-0.023}^{+0.021}$ $^{\mathrm{iii}}$ \\ 
$L_*$ [$L_{\odot}$] & $0.961_{-0.057}^{+0.058}$ & $0.948_{-0.070}^{+0.073}$ & - \\
$M_{V}$ [mag] & $4.877_{-0.072}^{+0.074}$ & $4.892_{-0.088}^{+0.090}$ & $4.77 \pm 0.09$ $^{\mathrm{iii}}$ \\
Age [Gyr] & $3.3_{-1.3}^{+1.9}$ & $3.1_{-1.6}^{+2.0}$ & $5.1_{-2.3}^{+2.7}$ $^{\mathrm{iii}}$ \\
Distance [pc] & $202.6_{-6.8}^{+6.8}$ & $201.2_{-8.2}^{+8.3}$ & $220 \pm 10$ $^{\mathrm{iii}}$ \\ 
$M_P$ [$M_J$] & $1.205_{-0.058}^{+0.058}$ & $1.202_{-0.051}^{+0.050}$ & $1.198 \pm 0.053$ $^{\mathrm{ii}}$ \\ 
$R_P$ [$R_J$] & $1.199_{-0.022}^{+0.020}$ & $1.187_{-0.035}^{+0.034}$ & $1.222 \pm 0.038$ $^{\mathrm{ii}}$ \\ 
$\rho_P$ [g\,cm$^{-3}$] & $0.870_{-0.041}^{+0.042}$ & $0.891_{-0.064}^{+0.074}$ & - \\ 
$a$ [AU] & $0.03566_{-0.00062}^{+0.00048}$ & $0.03563_{-0.00058}^{+0.00048}$ & $0.0367_{-0.0005}^{+0.0012}$ $^{\mathrm{i}}$\\ [1ex]
\hline\hline 
\end{tabular}
\label{tab:global} 
\end{table*}

\begin{figure}
\begin{center}
\includegraphics[width=8.4 cm]{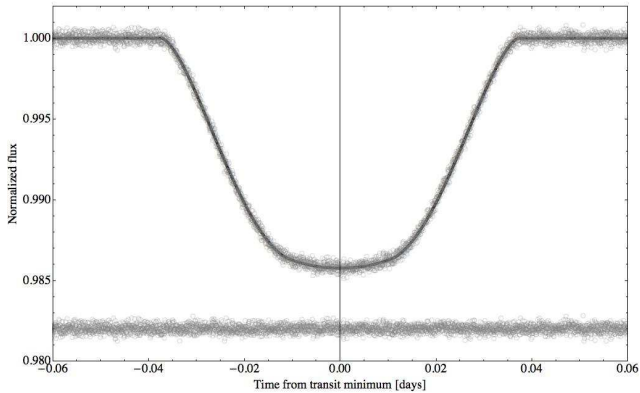}
\caption{\emph{Short-cadence folded transit lightcurve of TrES-2b (circles) with 
model fit overlaid. Residuals from the fit are shown below, offset by +0.982.}} 
\label{fig:globfit}
\end{center}
\end{figure}

\begin{figure*}
\begin{center}
\includegraphics[width=16.8 cm]{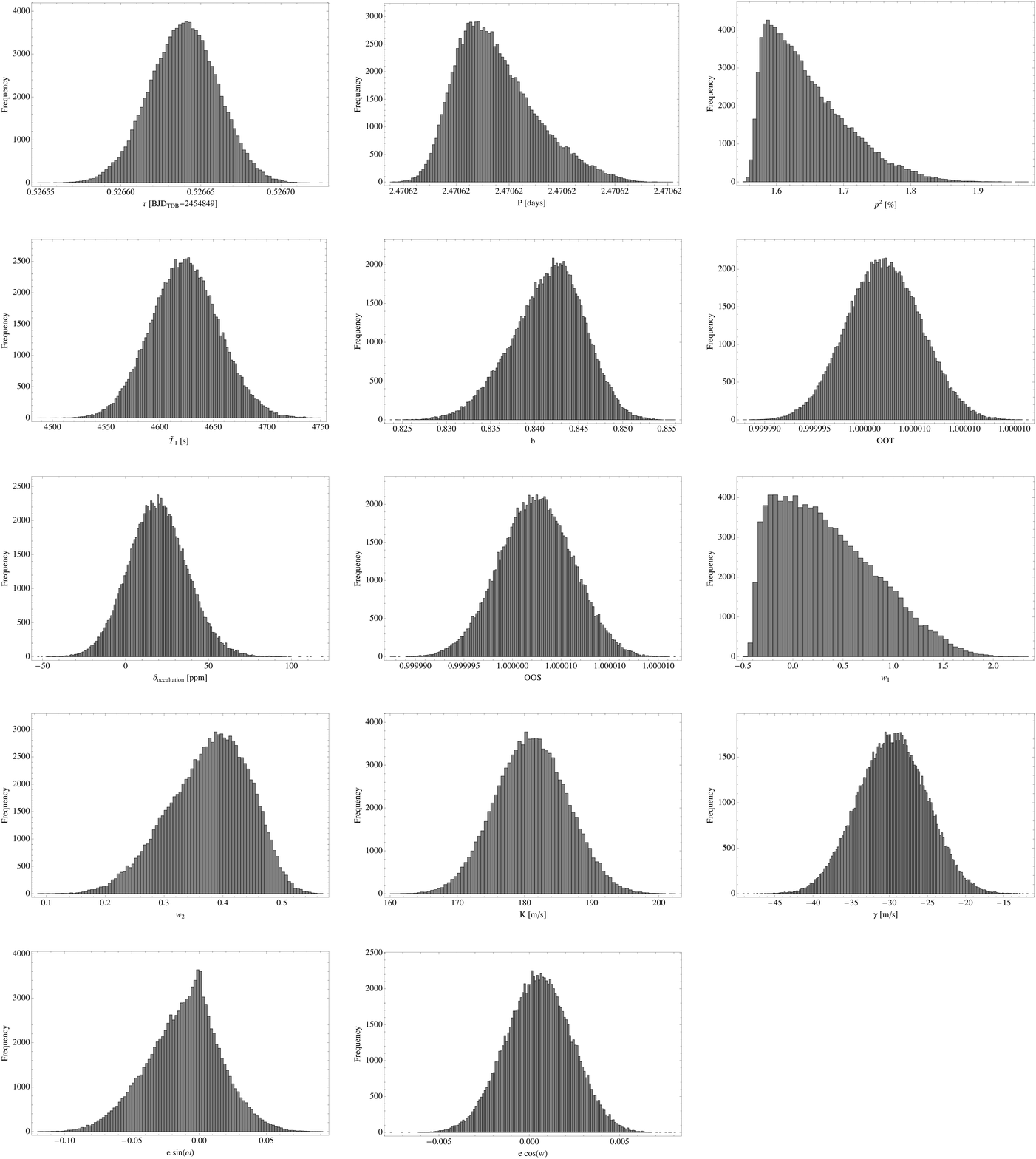}
\caption{\emph{Marginalized parameter posterior distributions. Reading down from
top left to right, we have $\tau$, $P$, $p^2$, $\tilde{T}_1$, $b$, OOT,
$\delta_{\mathrm{occultation}}$, OOS, $w_1$, $w_2$, $K$, $\gamma$, 
$e\sin\omega$ and $e\cos\omega$. Results come from global fits assuming
no Trojan body and no linear drift in the RVs, but allowing orbital 
eccentricity.}}
\label{fig:histos}
\end{center}
\end{figure*}

The global fits were performed using the full \emph{Kepler} time series as 
described in \S\ref{sub:modelgen}. The final results are given in 
Table~\ref{tab:global}. After the main MCMC fits, a downhill simplex routine
is used to obtain the lowest $\chi^2$ solution. We plot this solution over
the data in Figure~\ref{fig:globfit}\footnote{A high definition version of this 
figure is available at www.homepages.ucl.ac.uk/$\sim$ucapdki/globalfit.pdf}.
Histograms of the marginalized posterior distributions for each of the fitted
parameters are shown in Figure~\ref{fig:histos}, which clearly indicate
convergence of the fitting parameters.

\subsection{Limb Darkening Fitting}
\label{sub:ldfit}

\begin{figure}
\begin{center}
\includegraphics[width=8.4 cm]{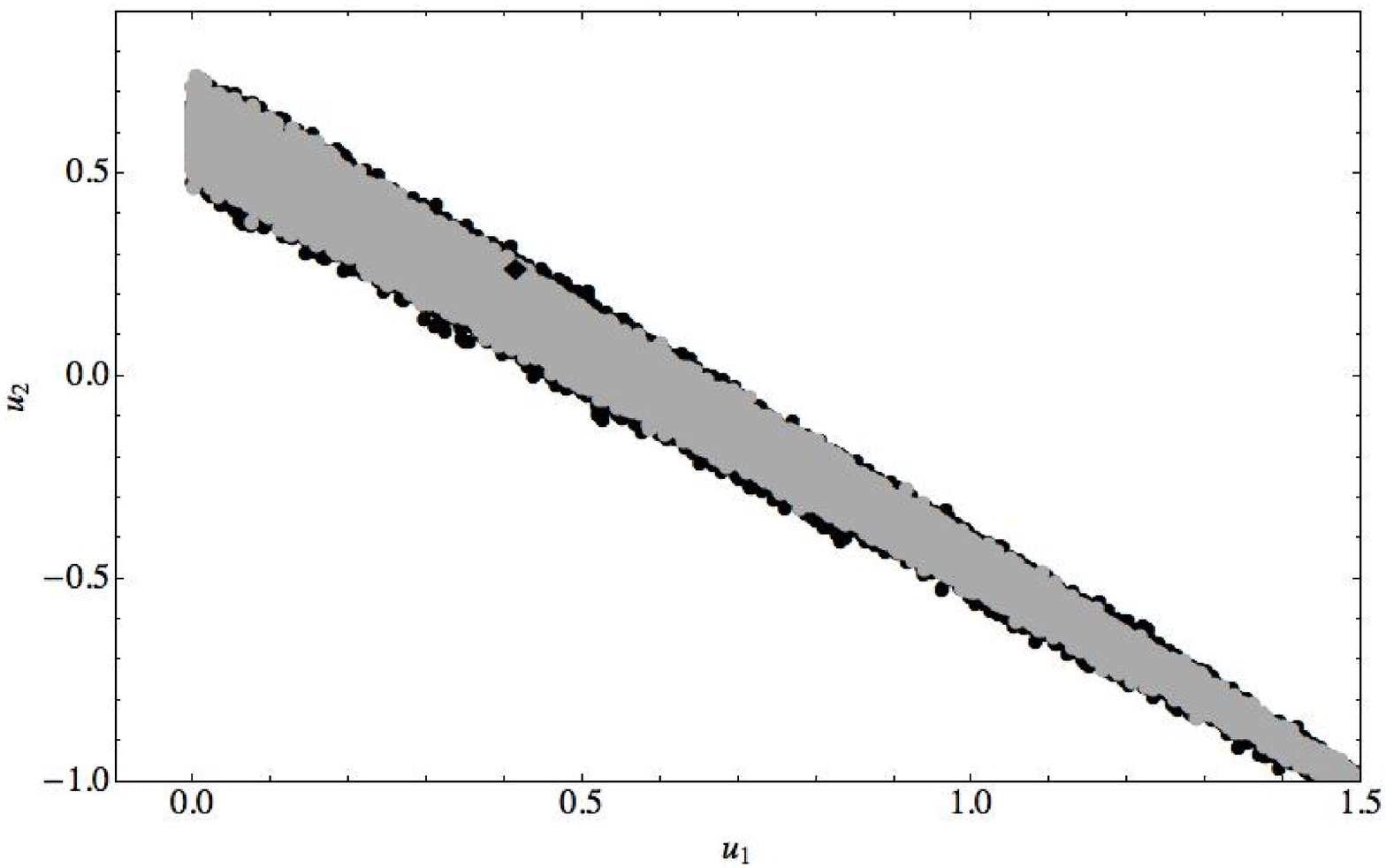}
\caption{\emph{Distribution of the quadratic limb darkening coefficients from 
the MCMC global fits of the short-cadence data. Black points correspond to the 
3-$\sigma$ region and gray to the 1-$\sigma$. The diamond marks the 
theoretical limb darkening coefficients computed from a \citet{kurucz:2006} 
style atmosphere.}} 
\label{fig:ldfit}
\end{center}
\end{figure}

Fitting for the limb darkening (LD) coefficients is challenging because TrES-2b 
is a near-grazing transit and thus only samples a fraction of the stellar 
surface. However, the extremely high quality of the \emph{Kepler} SC photometry 
and the fact we have 18 transits does allow for a good solution (given in 
Table~\ref{tab:global}). The inevitably strong correlations between the 
quadratic coefficients is presented in Figure~\ref{fig:ldfit}.

We find that the theoretical limb darkening coefficients lie within the 
1-$\sigma$ confidence region of our fits, indicating an impressive prediction 
for the \citet{kurucz:2006} atmosphere model. One major benefit of fitting for 
the limb darkening is that the uncertainty in the stellar properties is built 
into the model and thus leads to larger, and ultimately more realistic, 
estimates of the various parameter uncertainties. Parameters which are highly 
correlated to the limb darkening coefficients, such as the the transit depth 
(see \S\ref{subsub:whyld}), have their associated errors increase considerably 
as a result of this process.

As discussed in \S\ref{sub:fitting}, we actually fitted for $w_1$ and $w_2$
rather than $u_1$ and $u_2$ to decrease the correlations, following the
prescription of \citet{pal:2008}. We chose $\varphi=40^{\circ}$, as this
was suggested as a useful first-guess for the term by \citet{pal:2008}.
However, future studies of this system would benefit by using a more optimized
value of $\varphi$. By using a principal component analysis (PCA), we are able 
to find this optimum angle to be $\varphi = 42.7033^{\circ}$, very close to the 
$35^{\circ}-40^{\circ}$ range advocated by \citet{pal:2008}.

It is important to consider the effects of fitting for LD carefully. We re-ran 
our fits with the LD parameters fixed to their best-value and found that the 
errors on numerous parameters were considerably reduced, in many cases by an 
order-of-magnitude. As an example, the transit depth error is reduced by a 
factor of 17.5 when we fixed the LD parameters. The errors found using fitted 
LD correspond to the absolute uncertainty in each parameter. Therefore, if we 
wish to compare the duration found from \emph{Kepler} photometry with, say, a 
ground-based measurement in a different bandpass, we must fit for the LD 
parameters separately in both cases. However, if we consistently employ the same 
bandpass and instrument response function for the same star, then there is no 
need to refit the LD parameters everytime. By fixing the LD parameters to their 
best-value, we compute the relative duration changes, within that bandpass.

For TTV, the error in the time of transit minimum does not appreciably change 
between fitting and not-fitting the LD parameters. Therefore, the TTV seems to 
be reliable across different bandpasses and instruments. This opportunity will 
be exploited later in \S\ref{sec:longttv}.

\subsection{Occultation}
\label{sub:occultation}

In the short cadence global fits, we do not detect an occultation for the 
planet in the circular nor the eccentric fits. Although no occultation is
detected, a robust upper limit is obtained. We choose to use 
the eccentric fit from here on, as it provides the most realistic errors 
(see \S\ref{subsub:whyecc}).

The best-fitted occultation depth is 
$\delta_{\mathrm{occultation}} = 19_{-17}^{+18}$\,ppm, indicating no
detected signal. The posterior distribution of the occultation depth is 
presented in Figure~\ref{fig:histos}. We exclude an occultation of depth 
$>72.9$\,ppm to $3$-$\sigma$ confidence.

Recently, \citet{spiegel:2010} predicted that the occultation of TrES-2b, 
in \emph{Kepler}'s bandpass, would be $\leq 20$\,ppm, assuming no reflected 
light contribution. Our results are therefore highly consistent with the 
theoretical models for this planet.

The 3-$\sigma$ limit constrains the geometric albedo to be $A_g < 0.146$ and a 
dayside brightness temperature of $T_{P,day} < 2413$\,K (for comparison, the
equilibrium temperature is 1472\,K). We note that our 
3-$\sigma$ limit is tighter than that for HD 209458b as measured by 
\citet{rowe:2008} using MOST, where $A_g < 0.17$ to $3$-$\sigma$ confidence. 
Therefore, TrES-2b is currently the darkest exoplanet known to exist. For 
comparison, the upper limit corresponds to a planet of similar albedo to 
Mercury (0.138).

\subsection{A Search for Asymmetry}
\label{sub:asymmetry}

Lightcurve asymmetry is generally not expected but may reveal interesting, new
physics for hot-Jupiters. One possible source would be an oblate star with a
significant spin-orbit misalignment causing an asymmetry in the ingress/egress
durations. We here describe how we searched for asymmetry in the lightcurve.

\begin{figure}
\begin{center}
\includegraphics[width=8.4 cm]{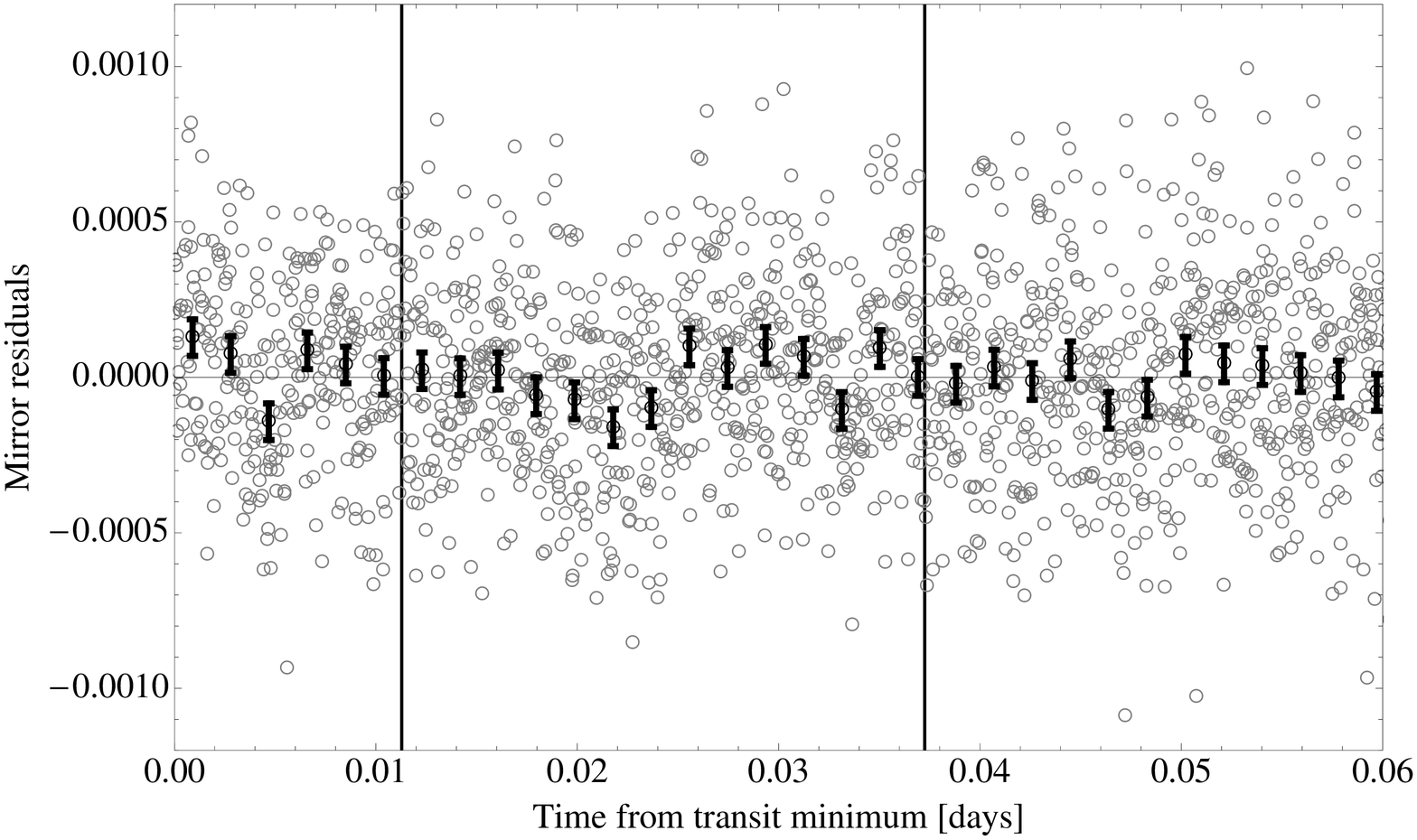}
\caption{\emph{Mirror residuals. Subtracting the two halves of the lightcurve
from each other produces the ``mirror'' residuals, which would reveal any
signs of lightcurve asymmetry - however, none are evident. The vertical black
lines mark the start of ingress/egress and the end of ingress/egress.}} 
\label{fig:mirror}
\end{center}
\end{figure}

We divide the folded lightcurve into points before and after the globally fitted 
time of transit minimum, where the fold is performed using the globally 
fitted period. We then mirror the two halves upon each other to search for signs 
of asymmetry in the lightcurve. The residuals of these two halves are shown
in Figure~\ref{fig:mirror}.

We first perform a linear interpolation of the folded data prior to the time of
transit minimum. This function is then evaluated at the time stamps of the 
folded data after the time of transit minimum and the associated uncertainties
are carried over. We then subtract the two and add the two sets of 
flux uncertainties in quadrature. The ``mirror residue'' exhibits an 
r.m.s. scatter of 305\,ppm, whereas from a theoretical point-of-view one 
expects scatter equal to $\sqrt{2} \times 237.2$\,ppm = $335$\,ppm. A chi 
squared test gives 6990 for 7684 data points. The ingress and egress therefore 
exhibit remarkable symmetry.

The most significant feature in Figure~\ref{fig:mirror} is a slight drop at
around +0.02\,days. This feature is only 2-$\sigma$ significant with the current
data, but could be scrutinized further in later data releases.

\subsection{Eccentricity}
\label{sub:eccentricity}

As shown in Table~\ref{tab:global}, we performed both circular and eccentric 
fits to illustrate the consequences of fitting for eccentricity. The fits find 
very similar $\chi^2$ values, with the circular fit being marginally larger by 
$\Delta\chi^2=1.89$ for 30697 data points. Using an F-test, we find that 
the eccentric fit is accepted over the circular fit with a confidence of 38.9\%, 
which we consider to be insignificant. Therefore, we conclude the orbit of 
TrES-2b is consistent with a circular orbit, based upon the current data. 
Further, using the marginalized posterior distribution, we estimate that the 
eccentricity satisfies $e<0.094$ to 3-$\sigma$ confidence.

The $e\cos\omega$ prior from \citet{donovan:2010} places a much stronger
constraint than that obtained for either component purely from the radial
velocity. As a result, we find a much larger 
uncertainty on $e\sin\omega$ than $e\cos\omega$. Whilst both components are 
consistent with zero, it is unlikely from an a-priori perspective than 
$e\sin\omega$ will be non-zero given that $e\cos\omega$ is essentially zero. If 
both components had similar uncertainties, but consistent with zero, then the 
eccentricity would be tied down to $e<0.0085$, but we stress that this is not a 
conclusion which can supported purely based upon the data.


\subsection{Revised Masses and Radii}
\label{sub:stellarparams}

Fundamental parameters of the host star such as the mass
($M_*$) and radius ($R_*$), which are needed to infer the planetary
properties, depend strongly on other stellar quantities that can be
derived spectroscopically.  For this we used the spectroscopic analysis
of \citet{sozzetti:2007} who determine $T_{\mathrm{eff}} = (5850 \pm 50)$\,K, 
[Fe/H]$=(-0.15 \pm 0.10)$ and $\log(g\,[\mathrm{cgs}])=(4.4 \pm 0.1)$.

In principle the effective temperature and metallicity, along with the
surface gravity taken as a luminosity indicator, could be used as
constraints to infer the stellar mass and radius by comparison with
stellar evolution models.

For planetary transits a stronger constraint is often provided by the
$a/R_*$ normalized semi-major axis, which is closely related to
$\rho_*$, the mean stellar density.  The quantity $a/R_*$ can be
derived directly from the transit lightcurve \citep{seager:2003} and the RV
data (for eccentric cases, see \citet{investigations:2010}). The results of our
100,000 MCMC trials are used to produce an array of 100,000 estimates for
$\rho_*$, $T_{\mathrm{eff}}$ and [Fe/H]. For every trial, we match stellar
evolution isochrones from \citet{yi:2001} to the observed properties to produce
100,000 estimates of the absolute dimensions of the star. Finally, the planetary 
parameters and their uncertainties were derived by the direct combination of the 
posterior distributions of the lightcurve, RV and stellar parameters.

After the first iteration for determining the stellar properties, as
described in \citet{bakos:2009}, we find that the surface gravity,
$\log(g\,[\mathrm{cgs}]) = 4.475_{-0.024}^{+0.024}$, is highly consistent with 
the \citet{sozzetti:2007} analysis.  Therefore, a second iteration (which would 
use the new $\log g$ value) of the isochrones was not required and we
adopted the values stated above as the final atmospheric properties of
the star (shown in Table~\ref{tab:global}).

The revised parameters are in excellent agreement with the estimates from 
\citet{donovan:2007}, \citet{holman:2007} and \citet{sozzetti:2007}, all shown 
in Table~\ref{tab:global} for comparison. However, our derived stellar density 
is markedly larger and this leads to a slightly smaller, more massive star, 
which consequently `deflates' TrES-2b slightly.

\section{TRANSIT TIMING VARIATIONS (TTV)}
\label{sec:TTV}

We will here only consider short-term transit timing variations, which we define 
to be those occurring within the timescales of the eighteen observed 
\emph{Kepler} transits in Q0 and Q1. A long-term transit timing analysis is 
provided in \S\ref{sec:longttv}.

\subsection{Fitting Method}
\label{sub:indivs}

For the individual fits, we do not expect limb darkening to vary from transit to 
transit and thus using a single, common set of LD coefficients is justified (as
explained earlier in \S\ref{subsub:whyLD}). We therefore fix the quadratic 
coefficients to those found to give the lowest $\chi^2$ in the global fit we 
performed earlier (selected values were $w_1 = 0.22366$ and $w_2 = 0.39288$). 
Aside from the limb darkening, the eccentricity terms $e\sin\omega$ and 
$e\cos\omega$ are also held fixed to the lowest $\chi^2$ solution 
($e\sin\omega=-0.020864$ and $e\cos\omega=0.00076088$). In total, there are five 
free parameters used: $\{\tau,p^2,\tilde{T}_1,b,\mathrm{OOT}\}$. An initial 
run is used to compute scaling factors individually for each transit epoch, 
which span $\pm0.125 P$ of the linear ephemeris predictions. The scaling factors 
are selected so that the lowest $\chi^2$ solution found is equal to the number 
of degrees of freedom in the model, as before.

\begin{figure*}
\begin{center}
\includegraphics[width=16.8 cm]{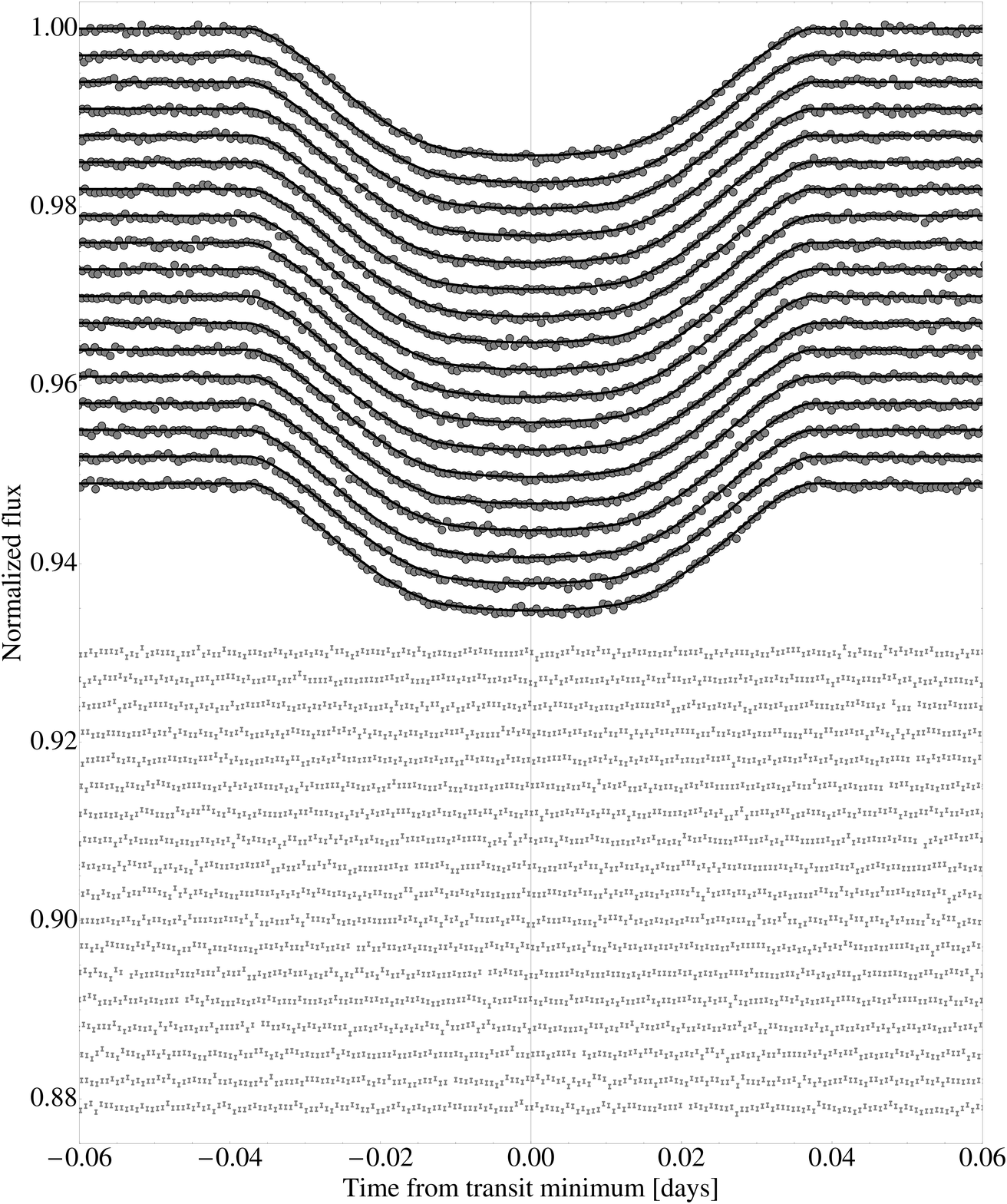}
\caption{\emph{Individual transits of TrES-2b from Q0 and Q1 \emph{Kepler}
photometry. Top lightcurve is Kepler epoch 0 going sequentially in
time to epoch 18 at the bottom. Residuals are shown below.}} 
\label{fig:indivs}
\end{center}
\end{figure*}

The individually fitted transit lightcurves are shown in Figure~\ref{fig:indivs} 
and the parameters in Table~\ref{tab:indiv}. We note that in none of the 
transits is a second transit-like feature observed, as claimed by 
\citet{raetz:2009}.

\subsection{Control Data}
\label{sub:control}

We describe here how we produce control data in the form of artificial
lightcurves. The act of producing control data by which to compare the
genuine observations is a practice frequently applied in many aspects of
scientific study. In our case, the control data serves two principal functions:

\begin{enumerate}
\item Rescaling of the parameter uncertainties
\item Identification of signals due to ``phasing''
\end{enumerate}

\subsubsection{Rescaling}
\label{subsub:rescaling}

Both of these issues were first noted in \citet{kipbak:2010}, although control
data was not used. The rescaling issue was observed by the authors as they found
that the errors produced by the Metropolis-Hastings MCMC method led to scatters 
in their TTV and TDV much lower than the parameter uncertainties. The chance of 
this occurring by coincidence in an isolated case was estimated to be 
$\sim$10\%, however the pattern was recurring for the majority of parameters 
evaluated. This led the authors to conclude that there was strong evidence the 
measurement uncertainties were being overestimated.

Calculating the necessary rescaling factor can be achieved by generating control
data. For a single global fit, as performed earlier, this would be too time
consuming with the 30,000+ data points plus the correlations between limb
darkening and depths taking several weeks to fit in a single run. Therefore,
the uncertainties presented in Table~\ref{tab:global} may in fact be 
overestimates as well, although we have not confirmed this. For the individual
data, one may take advantage of the fact that a planet exhibiting no TTV, TDV,
T$\delta$V (depth variations), TbV (impact parameter variations) or baseline
variations should yield a $\chi^2$ equal to the number of degrees of freedom in
each case. For example, for the TDV, we have 18 transits with one model
parameter, the mean duration, and so we expect $\chi^2=17$.

Since no real system can be assumed to be absolutely temporally invariant, the
only practical way forward is to generate artificial data for the control. To
accomplish this, we take the global fit model found earlier and sample it at the
exact time stamps in each individual transit epoch array. The global model
implicitly assumes that no parameters vary from epoch to epoch, satisfying our
control condition. Next, we introduce Gaussian noise into the lightcurve equal
to the actual noise recorded at those time stamps. This noise includes the
scaling factors found in the individual fits, to ensure equivalence. These
control lightcurves are then fitting using the Metropolis-Hastings MCMC method 
using identical starting positions, jump sizes, etc as the individual fits.

The various parameter variations are then evaluated and the necessary scaling
factor is computed. The scaling factors are given in the last line of 
Table~\ref{tab:indiv}. Our results agree with the conclusions of 
\citet{kipbak:2010} i.e. that all parameters have overestimated errors by around
a factor of $\sim2$. The reason for this overestimation is unclear and despite
close examination of our routines, we can find no obvious reason why this should
occur. An independent code used for HATNet discoveries \citep{bakos:2009}, 
which also uses Metropolis-Hastings MCMC, finds very similar uncertainties to 
the algorithm used in this work (detailed comparisons of the two methods have
been previously provided in \citet{kipbak:2010} and \citet{kipping:2011}), 
suggesting this is not a specific flaw in our routine. 

\subsubsection{Phasing}
\label{subsub:phasing}

In \citet{kipbak:2010}, the authors considered a new term which they labeled as 
the transit ``phasing''. This corresponds to the time difference between the 
expected time of transit minimum and the nearest data point. For example, for 
data of cadence 60\,s, we would expect this time difference to be in the range 
$\pm 30$\,s. Phasing does not seem have a linear correlation to observed
parameter variations, but does introduce false periods into the power spectrum
of the variations \citep{kipbak:2010}. Removing the phasing effects is not
currently possible, but we can at least generate the effects which phasing
induce to compare to the real data.

By generating our control data with the exact same cadence and time stamps
as the data which we fit in the individual transit arrays, we can recover any
possible influence the phasing may have on our results. In what follows,
figures showing parameter variations always have the real data on the 
left-hand-side and the control data on the right, so the effects of phasing
are most clearly visible.

\subsection{Analysis of Variance for TTV}
\label{sec:shortttv}

\begin{table*}
\caption{\emph{Transit parameters of TrES-2b from individual fits of the SC 
data. Kepler epoch 0 is defined as the first transit observed by \emph{Kepler}. 
In this data, we fix the limb darkening to the best fit values from the global 
fit. Therefore, the data can be used to look for relative changes between these 
18 transits, but not against previous ground-based measurements of TrES-2b. 
Errors have not been re-scaled, but the scaling factor is provided on the last 
line, based upon the analysis of control data.}} 
\centering 
\begin{tabular}{r c c c c c} 
\hline\hline 
Kepler Epoch & $\tau$ [BJD$_{\mathrm{TDB}}$ - 2,450,000] & $T_{1,4}$ [s] & $(R_P/R_*)^2$ [\%] & $b$ & OOT \\ [0.5ex] 
\hline
0 & $4955.763285_{-0.000096}^{+0.000096}$ & $6460_{-37}^{+37}$ & $1.634_{-0.016}^{+0.016}$ & $0.8454_{-0.0037}^{+0.0037}$ & $0.999995_{-0.000016}^{+0.000016}$ \\
1 & $4958.233958_{-0.000098}^{+0.000098}$ & $6447_{-39}^{+39}$ & $1.631_{-0.016}^{+0.017}$ & $0.8437_{-0.0042}^{+0.0040}$ & $0.999996_{-0.000016}^{+0.000017}$ \\
2 & $4960.704556_{-0.000095}^{+0.000095}$ & $6455_{-36}^{+36}$ & $1.628_{-0.015}^{+0.016}$ & $0.8444_{-0.0037}^{+0.0036}$ & $1.000014_{-0.000016}^{+0.000016}$ \\
3 & $4963.175188_{-0.000098}^{+0.000099}$ & $6426_{-39}^{+39}$ & $1.623_{-0.016}^{+0.016}$ & $0.8413_{-0.0041}^{+0.0040}$ & $1.000006_{-0.000022}^{+0.000022}$ \\
4 & $4965.645708_{-0.000094}^{+0.000095}$ & $6431_{-37}^{+37}$ & $1.624_{-0.015}^{+0.015}$ & $0.8401_{-0.0038 }^{+0.0037}$ & $1.000004_{-0.000016}^{+0.000016}$ \\
5 & $4968.116367_{-0.000093}^{+0.000093}$ & $6426_{-37}^{+37}$ & $1.624_{-0.015}^{+0.016}$ & $0.8397_{-0.0040}^{+0.0039}$ & $1.000015_{-0.000016}^{+0.000016}$ \\
6 & $4970.587001_{-0.000099}^{+0.000099}$ & $6434_{-39}^{+39}$ & $1.639_{-0.016}^{+0.017}$ & $0.8452_{- 0.0040}^{+0.0039}$ & $0.999998_{-0.000017}^{+0.000017}$ \\
7 & $4973.057600_{-0.000099}^{+0.000100}$ & $6469_{-38}^{+39}$ & $1.626_{-0.016}^{+0.016}$ & $0.8432_{-0.0039}^{+0.0037}$ & $1.000010_{-0.000017}^{+0.000017}$ \\
8 & $4975.528262_{-0.000103}^{+0.000102}$ & $6434_{-41}^{+41}$ & $1.621_{-0.017}^{+0.018}$ & $0.8400_{-0.0046}^{+0.0044}$ & $0.999992_{-0.000017}^{+0.000017}$ \\ 
9 & $4977.998830_{-0.000100}^{+0.000100}$ & $6430_{-40}^{+40}$ & $1.637_{-0.017}^{+0.017}$ & $0.8424_{-0.0043}^{+0.0041}$ & $1.000010_{-0.000017}^{+0.000017}$ \\
10 & $4980.469389_{-0.000099}^{+0.000098}$ & $6426_{-39}^{+40}$ & $1.615_{-0.016}^{+0.017}$ & $0.8407_{-0.0042}^{+0.0041}$ & $0.999989_{-0.000017}^{+0.000017}$ \\
11 & $4982.939972_{-0.000102}^{+0.000103}$ & $6504_{-41}^{+41}$ & $1.631_{-0.017}^{+0.017}$ & $0.8445_{-0.0041}^{+0.0040}$ & $0.999994_{-0.000017}^{+0.000017}$ \\
12 & $4985.410672_{-0.000097}^{+0.000096}$ & $6426_{-38}^{+38}$ & $1.634_{-0.016}^{+0.016}$ & $0.8434_{-0.0039}^{+0.0038}$ & $0.999991_{-0.000016}^{+0.000016}$ \\
13 & $4987.881247_{-0.000096}^{+0.000097}$ & $6403_{-37}^{+37}$ & $1.626_{-0.015}^{+0.016}$ & $0.8403_{-0.0039}^{+0.0037}$ & $0.999991_{-0.000016}^{+0.000016}$ \\
14 & $4990.351850_{-0.000098}^{+0.000098}$ & $6451_{-39}^{+39}$ & $1.631_{-0.017}^{+0.017}$ & $0.8427_{-0.0041}^{+0.0040}$ & $1.000001_{-0.000016}^{+0.000017}$ \\
15 & $4992.822571_{-0.000098}^{+0.000097}$ & $6448_{-39}^{+39}$ & $1.629_{-0.016}^{+0.017}$ & $0.8404_{-0.0042}^{+0.0040}$ & $0.999997_{-0.000017}^{+0.000017}$ \\
16 & $4995.293071_{-0.000098}^{+0.000099}$ & $6462_{-39}^{+39}$ & $1.629_{-0.016}^{+0.016}$ & $0.8436_{-0.0040}^{+0.0039}$ & $1.000016_{-0.000016}^{+0.000016}$ \\
17 & $4997.763671_{-0.000103}^{+0.000102}$ & $6439_{-40}^{+40}$ & $1.623_{-0.017}^{+0.017}$ & $0.8398_{-0.0043}^{+0.0042}$ & $0.999999_{-0.000019}^{+0.000018}$ \\ 
\hline
Scaling Factor & 0.5695 & 0.6036 & 0.5044 & 0.5574 & 0.5171 \\ [1ex]
\hline\hline 
\end{tabular}
\label{tab:indiv} 
\end{table*}

\begin{figure*}
\begin{center}
\includegraphics[width=16.8 cm]{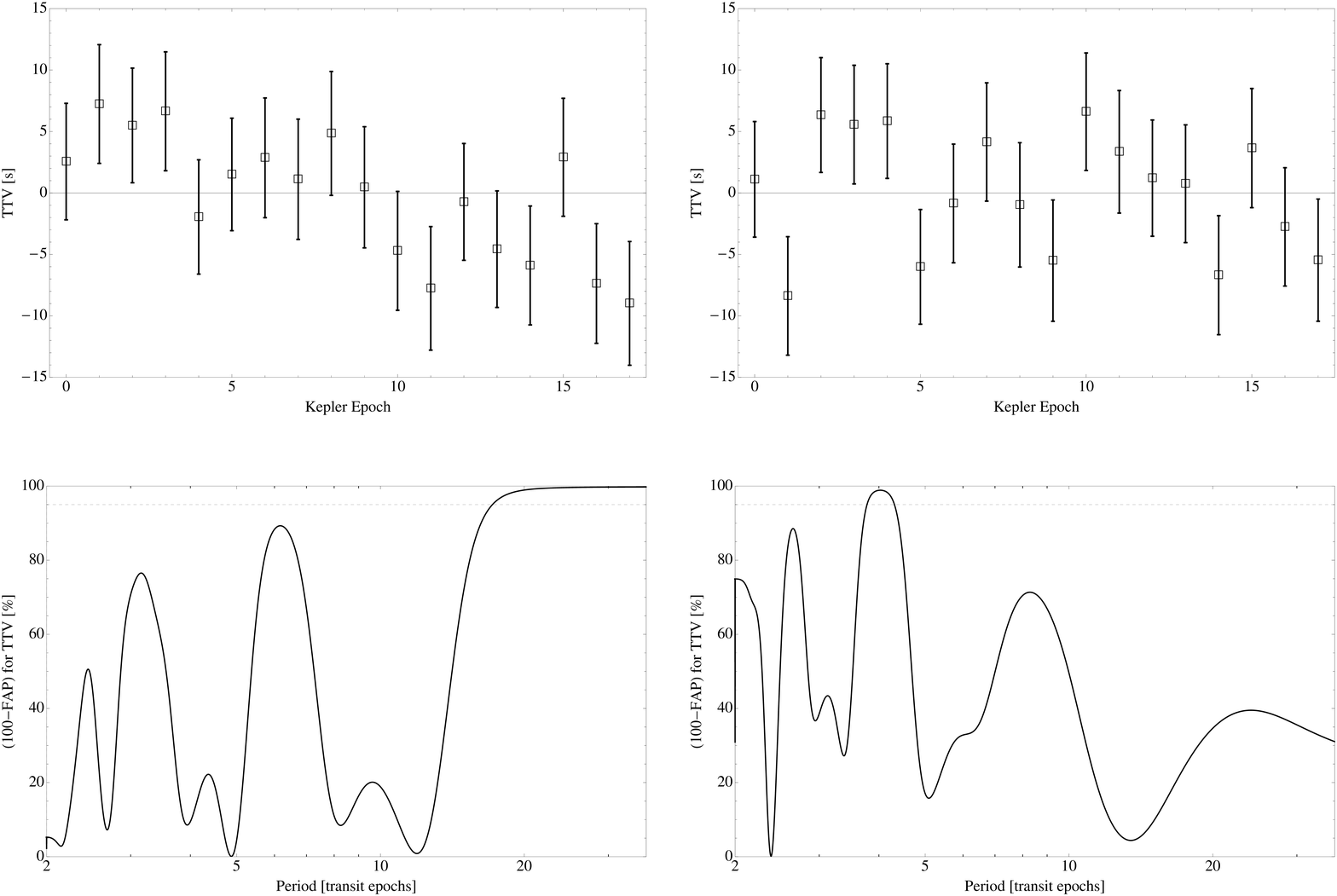}
\caption{\emph{TTVs of TrES-2b. Top left: Observed TTVs for TrES-2b using the
ephemeris of the global fit. Bottom left: F-test periodogram of the observed
TTVs. Top right: TTVs computed from control data (artificial lightcurves).
Bottom right: F-test periodogram of the control TTVs.}} 
\label{fig:ttvs}
\end{center}
\end{figure*}

The TTV, shown in the top-left panel of Figure~\ref{fig:ttvs}, exhibits a 
r.m.s. scatter of 5.16\,s, which demonstrates the impressive precision of 
these \emph{Kepler} measurements. After rescaling the uncertainties, the scatter 
in the data is consistent with a linear ephemeris, exhibiting a $\chi^2 = 19.0$ 
for 16 degrees of freedom. The excess scatter is 1.1-$\sigma$ significant, which
we consider below our detection threshold. The unscaled errors yield 
$\chi^2 =6.2$ for 16 degrees of freedom, supporting the hypothesis that the 
errors are significantly underestimated and justifying our rescaling 
methodology. Figure~\ref{fig:ttvs} shows the results.

\subsection{F-test Periodogram for TTV}
\label{sub:ttvperiodogram}

The F-test periodogram fits sinusoidal waveforms through the data of 
various periods, stepping through from the Nyquist frequency to the 
observational window in equally spaced steps of size 1/1000 of an epoch. Fitting 
for amplitude and phase, the $\chi^2$ is computed in each step, and then the 
F-test is performed. The false-alarm-probabilities (FAP) of these F-tests are 
then plotted in a periodogram. It is important to appreciate that the F-test is 
designed to look for sinusoidal waveforms, and thus periodic but non-sinusoidal 
waveforms would have their significances attenuated.

The control data reveals periodogram peaks at 2, 4 and 8 cycles which are
harmonics of the sampling cadence of one transit measurement per transit epoch.
In the real data, only one peak surpasses 95\% confidence occurring at a period
longer than the observation window. Such long period peaks cannot be considered
genuine unless further transit epochs confirm the periodicity. In conclusion,
there is no evidence for a TTV signal in the \emph{Kepler} Q0 and Q1 TrES-2b 
photometry.

\subsection{Excluded TTV Signals}
\label{sub:excludedttvs}

We conclude our analysis of the TTV by evaluating the constraints on other 
planets, moons and Trojans in the system. For 16 degrees of freedom, r.m.s. 
scatter producing a $\chi^2 = 36.2$ is excluded to 3-$\sigma$ confidence. 
This excludes r.m.s. scatter of 7.11\,s to the same confidence level.

An outer perturbing planet in a $j$:$j+1$ mean motion resonance (MMR) would 
cause the inner transiting planet to librate leading to TTVs 
\citep{holman:2005,agol:2005}. For 1:2, 1:3 and 1:4 resonances, the 
libration periods are 18.1, 10.5 and 7.2 cycles respectively. We therefore 
possess sufficient baseline to look for all such resonant planets. This excludes 
the presence of coplanar, MMR planets in these resonances of 0.11\,$M_{\oplus}$, 
0.17\,$M_{\oplus}$ and 0.22\,$M_{\oplus}$ respectively.

For an extrasolar moon in a retrograde orbit, the maximum dynamically stable 
orbital separation is 0.9309\,Hill radii \citep{domingos:2006}. For such a body 
on a circular, coaligned orbit, we are able to exclude moons of 
1.15\,$M_{\oplus}$. As the orbital separation decreases, we are able to exclude 
moons of masses $\geq (1.07/f)\sin i_s$\,$M_{\oplus}$, where $f$ is equal to the 
moon's orbital separation in units of Hill radii.

Trojan bodies can also induce TTVs and thus constraints on their presence
can be also established. Using Equation~1 of \citet{ford:2007}, and assuming a 
Trojan of angular displacement $\sim10^{\circ}$ from the Lagrange point, we are 
sensitive to Trojans of cumulative mass $>0.46$\,$M_{\oplus}$ to 3-$\sigma$ 
confidence. However, the expected libration period would be $\sim75$\,cycles and 
thus we do not yet possess sufficient baseline to definitively exclude such 
bodies.

\subsection{Proposed 0.21\,Cycle Period Signal}
\label{sub:rabussignal}

Another signal we are able to investigate is the one proposed by 
\citet{rabus:2009}. The authors claimed a 0.21 cycle period sinusoid of 
amplitude 50\,s provided a best-fit to the previously known transit times of 
TrES-2b, with a FAP of 1.1\% and suggested a 52\,$M_{\oplus}$ exomoon as a 
possible origin. 

Fixing the amplitude and period to the proposed value and fitting for the phase 
term, we find a $\chi^2 = 1610$ for the 18 data points. In contrast, the 
static model obtains a $\chi^2 = 19.0$, which therefore excludes the claimed
signal to high confidence. This highlights the dangers of looking for signals 
below the Nyquist frequency.

\section{TRANSIT DURATION VARIATIONS (TDV)}
\label{sec:TDV}

\subsection{Choosing a Statistic}
\label{sub:tdvstat1}

Due to the near-grazing nature of the transit, the standard assumption that 
$\tilde{T}$ is the optimum statistic for TDV searches may not be valid 
\citep{carter:2008,kipping:2009}. In particular, the first-to-fourth 
contact duration, $T_{1,4}$, could potentially offer greater sensitivity. We 
found that the typical error on $T_{1,4}$ was 0.28\% in the individual fits, 
whereas $\tilde{T}_1$ marginally better at 0.24\%. Another factor in choosing
a statistic comes from the effects of limb darkening. Whilst here we fix the
limb darkening coefficients to the best-fit values from the global MCMC, it
is preferable to still avoid using a statistic which is strongly correlated to
limb darkening. This is because we may be using slightly incorrect limb
darkening coefficients which therefore feed into incorrect duration estimations.
Whilst this is generally unavoidable, strongly correlated terms would clearly
excerbate the situation. We find that $T_{1,4}$ has a correlation of 0.021 to
the limb darkening coefficient $w_2$ (the most strongly constrained coefficient)
whereas $\tilde{T}_1$ has a correlation coefficient of 0.81. On this basis,
we choose to use the $T_{1,4}$ statistic in what follows, defining the
TDV of the $i^{\mathrm{th}}$ transit measurement as:

\begin{align}
\mathrm{TDV}_i = [T_{1,4}]_i - [T_{1,4}]_{\mathrm{global}}
\label{eqn:TDVdefn}
\end{align}

\subsection{Analysis of Variance for TDV}
\label{sub:shorttdv}

\begin{figure*}
\begin{center}
\includegraphics[width=16.8 cm]{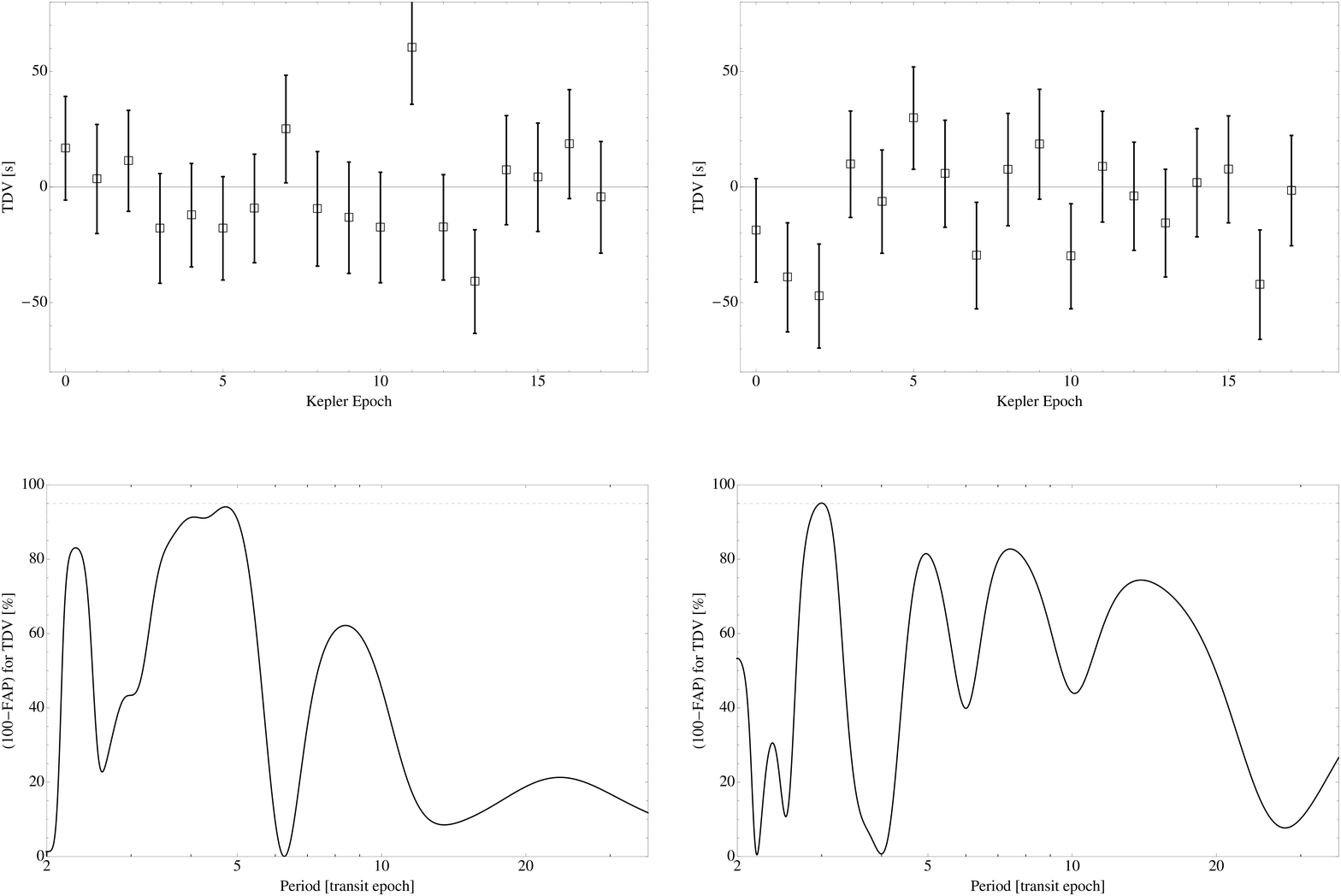}
\caption{\emph{TDVs of TrES-2b. Top left: Observed TDVs for TrES-2b using the
duration of the global fit. Bottom left: F-test periodogram of the observed
TDVs. Top right: TDVs computed from control data (artificial lightcurves).
Bottom right: F-test periodogram of the control TDVs.}} 
\label{fig:tdvs}
\end{center}
\end{figure*}

The TDV, shown in the top-left panel of Figure~\ref{fig:tdvs}, exhibits a 
r.m.s. scatter of 22.4\,s. After rescaling the uncertainties, the scatter in 
the data is consistent with a constant duration, exhibiting a $\chi^2 = 15.4$ 
for 17 degrees of freedom. The unscaled errors yield 
$\chi^2 =5.6$ for 17 degrees of freedom, again supporting the hypothesis that 
the errors are significantly underestimated. Figure~\ref{fig:tdvs} shows the 
results.

\subsection{F-test Periodogram for TDV}
\label{sub:tdvperiodogram}

We continue by computing the F-test periodogram for the TDV data (shown in 
lower-left of Figure~\ref{fig:tdvs}). The TDV data yields only one interesting
peak occurring with a broad distribution surrounding $4.72 \pm 0.10$\,cycles, 
significance 94.1\%. Firstly, this is below our formal detection threshold.
Secondly, the peak seems to occur in the control data, with distinct phasing
periods occurring at 3, 5 and 8 cycles. In light of this, we do not consider
the signal to be genuine.

\subsection{Excluded TDV Signals}
\label{sub:excludedtdvs}

The TDVs exclude a signal of r.m.s. amplitude 35.1\,s to 3-$\sigma$ confidence, 
or variations in the duration of 0.77\% over the 18 cycles. This excludes 
exomoons inducing TDV-V of mass $\geq 23.5 \sqrt{f} \cos i_s$\,$M_{\oplus}$ to 
the same confidence level. Additionally, it excludes 
$f M_S \sin i_s \geq 1.17$\,$M_{\oplus}$ through the TIP effect. 

\begin{figure*}
\begin{center}
\includegraphics[width=16.8 cm]{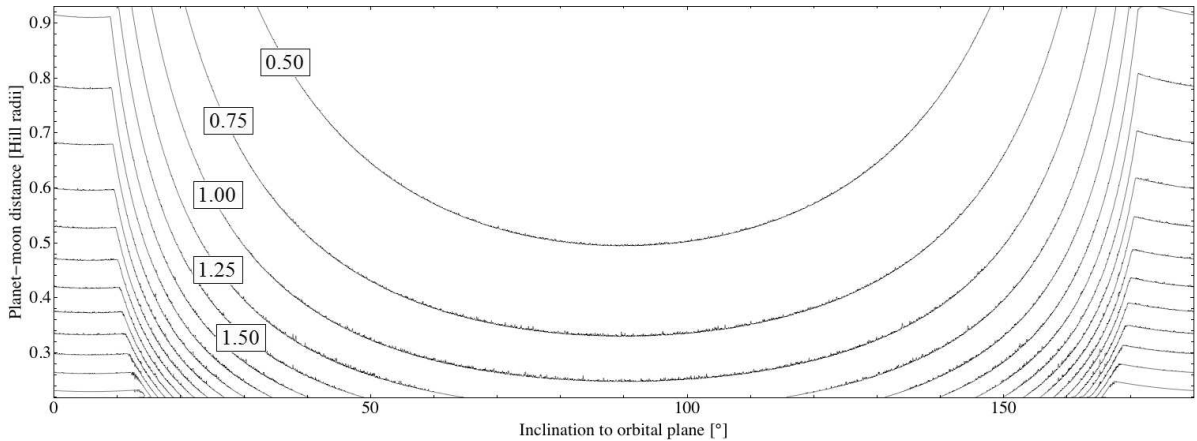}
\caption{\emph{Excluded exomoon masses for a companion to TrES-2b, as a function 
of the orbital distance of the moon around TrES-2b and the orbital inclination 
with respect to the orbital plane of TrES-2b. Contours are given in units of 
Earth masses, making steps of 0.25\,$M_{\oplus}$. The \emph{Kepler} data is able 
to easily probe down to sub-Earth mass exomoons.}} 
\label{fig:moonlimits}
\end{center}
\end{figure*}

Combining the TTV limits, the TDV-V limits and the TDV-TIP limits allows us to 
plot the parameter space of excluded exomoon masses, at the 3-$\sigma$ 
confidence level, assuming a circular orbit in Figure~\ref{fig:moonlimits}. We
make use of the expressions for the TTV, TDV-V and TDV-TIP presneted in
\citet{kipping:2009a,kipping:2009b}. We find that \emph{Kepler} is clearly 
sensitive to sub-Earth mass exomoons, as predicted by \citet{kipping:2009}.

Figure~\ref{fig:moonlimits} shows that for moons co-aligned to the planet's
orbital plane ($i_S = 90^{\circ}$), moons down to sub-Earth mass are excluded.
The sensitivity drops off as inclination increases away from a co-aligned
system but stabilizes for highly inclined moons (the kinks close 
$i_S\sim0^{\circ}$ and $i_S\sim180^{\circ}$) as a result of TDV-TIP effects
dominating.

\subsection{Proposed Inclination Change}
\label{sub:incchange1}

\citet{mislis:2009} claimed to have detected a linear decrease in the duration 
of TrES-2b due to the inclination angle varying at a rate of $-0.195^{\circ}$ 
over $\sim$300\,cycles, or $-0.00065^{\circ}$ per cycle.

Because the other ground-based measurements did not have their limb darkening 
coefficients fitted for, a fair comparison is not possible, in our view. 
Although the expression for the duration is independent of limb darkening 
parameters, we found that the duration was highly correlated to the limb 
darkening coefficients. However, we are able to use our 18 measurements of the 
inclination to quantify the constraints on the rate of inclination change in 
this system. Comparing data taken from the same instrument which has a constant 
CCD response function and bandpass is justifiable without fitting for limb 
darkening each time, since the LD parameters will not vary transit-to-transit.

Fitting a linear trend through our inclination data gives a rate of change of 
$+(0.0019 \pm 0.0020)^{\circ}$ per cycle, which is clearly not significant. 
We exclude an inclination change of $-0.0041^{\circ}$ per cycle to 3-$\sigma$ 
confidence, which is larger than that claimed by \citet{mislis:2009}. Therefore, 
using the current \emph{Kepler} data alone is not sufficient to yet
confirm/reject the proposed inclination change in this system, largely due to 
the very small temporal baseline of just 18 cycles. We will return to this 
hypothesis in our study on the long term timing changes in \S\ref{sub:longtdv}.

\subsection{Other Changes}

\subsubsection{Baseline}

The baseline fluxes are in excellent agreement with the global mean at all
epochs, giving $\chi^2 = 18.8$ for 16 degrees of freedom. This is not
a surprise since the baseline has been normalized twice during our corrective
procedure (see \S\ref{sub:detrending}) to ensure precisely this result.

\subsubsection{Transit depth}

\begin{figure}
\begin{center}
\includegraphics[width=8.4 cm]{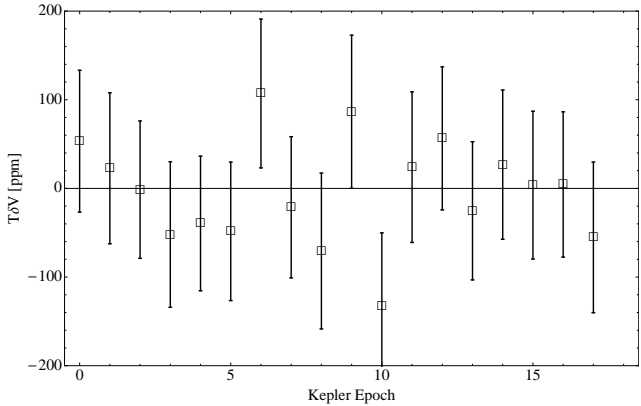}
\caption{\emph{Transit depth variations (T$\delta$V) of TrES-2b.}} 
\label{fig:depthchange}
\end{center}
\end{figure}

The transit depths are extremely stable yielding $\chi^2=8.7$ for 16 degrees
of freedom, which is our most stable statistic. The T$\delta$Vs (transit depth
variations) are shown in Figure~\ref{fig:depthchange}, where the low scatter,
of standard deviation 59.3\,ppm, is evident. We exclude variations of 123\,ppm
to 3-$\sigma$ confidence. Over the timescale of years,
transiting planets on periods $>10$\,days may exhibit variations due to
precession of an oblate planet's rotation axis \citep{carterwinn:2010}.
However, we do not possess sufficient baseline to look for such effects with
the 18\,cycles of this study.

\section{LONG-TERM TIMING VARIATIONS}
\label{sec:longttv}

\subsection{Ephemeris Fitting}
\label{sub:ephemeris}

In Table~\ref{tab:allttv} of the appendix, we show all of the measurements of 
the transit times of TrES-2b used in this study, including both amateur and 
professional measurements. 
The inclusion of the previous transit times leads to much tighter constraints on 
the period and epoch. We repeated our fits without using the previous transit 
times and found a local period of $P=2.4706112_{-0.0000018}^{+0.0000024}$\,days. 
Using all of the transit times yields 
$P=2.47061892_{-0.00000012}^{+0.00000018}$\,days, which is slightly longer
than that found using the \emph{Kepler} data only (note the much higher
precision of using all of the transit times). This discrepancy is visible
in Figure~\ref{fig:ttvs} where a drift in the TTVs is apparent and an excess of
low-frequency power exists in the periodogram. Whilst both values are
consistent with the \citet{holman:2007} values of $P=2.470621\pm0.000017$\,days,
the reason for this discrepancy warrants further investigation.

The previous transit times have several differences to the \emph{Kepler} times;
they are mostly from amateur astronomers and they have a longer 
temporal baseline by a factor of 31.5. Another difference is that the
DAWG do not recommend basing scientific conclusions on the \emph{Kepler} times
to an absolute accuracy of less than 6.5\,s, until such a time as this level
of accuracy can be verified (see \label{sub:timestamps}). 6.5\,s is certainly
sufficient to explain the observed low-frequency power observed in 
Figure~\ref{fig:ttvs}. We therefore consider three possible hypotheses to
explain the discrepancy:

\begin{enumerate}
\item The amateur transit times are unreliable and bias our results
\item There exists a long-term deviation away from a linear ephemeris
\item Systematic error in the \emph{Kepler} times
\end{enumerate}

\subsubsection{Hypothesis 1 - The ETD measurements are unreliable}

The list of transit times used in this study consists of 62 amateur measurements
from the ETD and 22 from peer-reviewed publications. The professional times
should be considered reliable by virtue of the peer review process, however
the amateur times may or may not be reliable. For several epochs, there
are simultaneous measurements from both camps and one may use these to evaluate
the reliability of the amateur measurements. 

Epoch 142 has two measurements from each camp (a total of four transit times). 
The weighted average of the professional measurements from 
\citet{rabus:2009} and \citet{raetz:2009} average to 
$\mathrm{BJD}_{\mathrm{TDB}}$\,$2454308.46163$. Each timing measurement deviates
from this point by $0.76$-$\sigma$ and $2.18$-$\sigma$ for the 
\citet{rabus:2009} and \citet{raetz:2009} measurements respectively. The ETD
amateur measurements deviate from this same point by 0.13-$\sigma$ and 
0.76-$\sigma$, indicative of a highly consistent result.

Epoch 278 has one from each camp, the professional measurement being from 
\citet{raetz:2009}. The amateur measurement deviates from this point by 
0.93-$\sigma$, even when negating the error on the professional measurement 
(0.74-$\sigma$ when including both errors).

Epoch 316 has one from each camp, the professional measurement being from 
\citet{raetz:2009}. The amateur measurement deviates from this point by 
1.89-$\sigma$, even when negating the error on the professional measurement 
(0.78-$\sigma$ when including both errors).

Epoch 395 has one from each camp, the professional measurement being from 
\citet{mislis:2010}. The amateur measurement deviates from this point by 
0.083-$\sigma$, even when negating the error on the professional measurement 
(0.076-$\sigma$ when including both errors).

Finally, epoch 414 is contemporaneous with one our \emph{Kepler} transits 
(Kepler epoch 10), three ETD measurements and one professional time
from \citet{mislis:2010}. Neglecting the much smaller error on our \emph{Kepler}
transit time, the \citet{mislis:2010} time deviates away by 1.88-$\sigma$. The
ETD measurements deviate by 3.18-$\sigma$, 0.43-$\sigma$ and 0.22-$\sigma$.

In conclusion, the contemporaneous sample of eight amateur transit times indicates 
that the amateur measurements are highly consistent with both the professional 
data and the \emph{Kepler} times. Seven out of eight measurements were less than 
1-$\sigma$ away from the professional determination, with one outlier at 
$3$-$\sigma$. This outlier point would be disregarded automatically by our 
fitting algorithm anyway as a result of using median statistics (see 
\S\ref{subsub:transittimes}). Although we cannot perform this test on every
signle amateur transit, it seems reasonable that the selected sample is a 
fair representation of the ETD database.

\subsubsection{Hypothesis 2 - Long-term non-linear ephemeris}

Having shown that hypothesis 1 is not supported by the current body of evidence,
we move on to investigate our second hypothesis. We tried fitting all of the 
data again using a quadratic ephemeris through the transit times, using the same
approach as that of \citet{holman:2010}. The model is 
$\tau_N = \tau_0 + n P + n^2 C$, where $n$ is the epoch number and $C$ is the 
curl. For a period which decreases over time, one expects $C<0$.

\begin{figure}
\begin{center}
\includegraphics[width=8.4 cm]{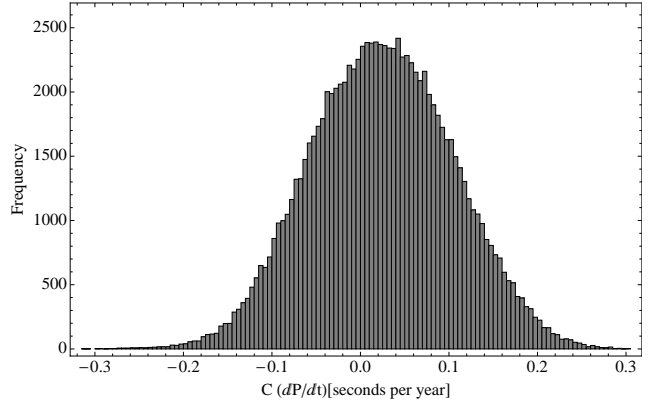}
\caption{\emph{Marginalized posterior distribution of the rate of change of
the orbital period for TrES-2b ($C$). We find no evidence for a long-term change
in the planet's orbital period.}} 
\label{fig:curl}
\end{center}
\end{figure}

Using both the $\chi^2$ and $\xi^2$ statistics, we found that there is no
preference for a quadratic trend in the transit times. In Figure~\ref{fig:curl},
we show the marginalized posterior distribution of $C$, which is symmetric about
zero. The data excludes a change in the orbital period of $0.11$\,seconds per
year, to 3-$\sigma$ confidence.

\subsubsection{Hypothesis 3 - Systematic error in the \emph{Kepler} times}

The only contemporaneous transit measurement between \emph{Kepler} and the
professional measurements is insufficient to test this hypothesis. This is
because the transit is question, measured by \citet{mislis:2010} for epoch 414,
has a precision of 518\,s and is therefore not useful for testing 
\emph{Kepler}'s timing accuracy at the $<6.5$\,s level.

In conclusion, the current body of evidence is insufficient to determine the
cause of the discrepant periods. However, further transits from \emph{Kepler}
will resolve this issue in the future.

\subsection{Duration Change}
\label{sub:longtdv}

\subsubsection{Comparing different bandpasses}

As discussed in \S\ref{sub:incchange1}, \citet{mislis:2009} claimed to have 
detected a linear decrease in the duration of TrES-2b due to the inclination 
angle varying at a rate of $0.195^{\circ}$ over $\sim$300\,cycles, or 
$0.00065^{\circ}$ per cycle.

In order to look for evidence of long duration change, it is necessary to use 
data taken before the \emph{Kepler Mission}. \citet{holman:2007} obtained three 
high-quality transits observations using the FLWO 1.2\,m telescope, in 
anticipation of this requirement, with a mean cycle value of $20.\dot{6}$. In 
contrast, the Q0 and Q1 data have a mean cycle value of 412.5, giving a baseline 
to the FLWO data of $391.8\dot{3}$\,cycles.

However, as also discussed in \S\ref{sub:incchange1}, we cannot compare data 
from different bandpasses unless we fit for limb darkening coefficients, 
especially since transit parameters are known to be acutely correlated to limb 
darkening for near-grazing transits (see \S\ref{subsub:whyld} and 
\S\ref{sub:ldfit}). \citet{holman:2007} did not fit for limb darkening 
coefficients and so we here present a re-analysis of those three transits. 

\subsubsection{Re-analysis of \citet{holman:2007} photometry}

We choose to fit for linear limb darkening due to the lower signal-to-noise from 
ground-based data. Using the corrected photometry from \citet{holman:2007}, we 
perform these fits in the same manner used in this paper. We float $e\sin\omega$, 
$e\cos\omega$, $K$, $\gamma$ and $P$ around their best-fit values from the 
global eccentric run (see Table~\ref{tab:global}) to allow their errors to 
propagate into the MCMC. The results are reported in Table~\ref{tab:holman}.

\subsubsection{Choosing a statistic}

In \S\ref{sub:tdvstat1}, we compared durations within the same bandpass and thus 
fixing limb darkening was justified. Since $\tilde{T}_1$ is highly correlated to 
the limb darkening coefficients, we find $T_{1,4}$ offers the highest 
signal-to-noise for TrES-2b when limb darkening is fitted and will be adopted 
here.

\subsubsection{Limits on duration change}

The \citet{holman:2007} global fit finds $T_{1,4} = 6559_{-103}^{+102}$\,s
whereas the \emph{Kepler} global fit finds $T_{1,4} = 6439_{-28}^{+25}$\,s,
giving $\Delta T_{1,4} = (-120 \pm 106)$\,s over $391.8\dot{3}$\,cycles, which 
we do not consider to be significant. The data 
exclude a decrease in the transit duration $|\Delta T_{1,4}| > 438$\,s, or
165\,s per year, to 3-$\sigma$ confidence. In contrast, \citet{mislis:2009} 
claim to have detected a duration decrease of $\sim3.16$\,minutes (189.6\,s) 
over $\sim300$\,cycles (or 91\,s per year).
Whilst this is not supported by our analysis, it is also not excluded.
\citet{scuderi:2010} have challenged the \citet{mislis:2009} result recentlty
using ground-based transit observations and the \citet{gilliland:2010} result. 
Future \emph{Kepler} transits will resolve this issue definitively.

\begin{table}
\caption{\emph{Results from re-analysis of TrES-2b using the \citet{holman:2007} 
photometry. Quoted values are medians of MCMC trials with errors given by 
1-$\sigma$ quantiles. * = fixed parameter; $\dagger$ = parameter was floated but 
not fitted.}} 
\centering 
\begin{tabular}{c c c} 
\hline\hline 
Parameter & Our fit & H07 value \\ [0.5ex] 
\hline 
$P$ [days] & $2.47061892_{-0.00000012}^{+0.00000018}$ $\dagger$ & $2.470621 \pm 0.000017$ \\
$T_{1,4}$ [s] & $6559_{-103}^{+102}$ & $6624 \pm 72$ \\
$\tilde{T}_1$ [s] & $4656_{-196}^{+155.6}$ & - \\
$T_{2,3}$ [s] & $1706_{-913}^{+505}$ & - \\
$(T_{1,2} \simeq T_{3,4})$ [s] & $2432_{-229}^{+411}$ & $2459 \pm 162$ \\
$(R_P/R_*)^2$ [\%] & $1.674_{-0.118}^{+0.052}$ & - \\
$b$ & $0.848_{-0.018}^{+0.022}$ & $0.8540 \pm 0.0062$ \\
$e\sin \omega$ & $-0.009_{-0.029}^{+0.024}$ $\dagger$ & $0^{*}$ \\
$e\cos \omega$ & $0.0005_{-0.0018}^{+0.0018}$ $\dagger$ & $0^{*}$ \\
$\Psi$ & $0.973_{-0.082}^{+0.071}$ $\dagger$ & $1^{*}$ \\
$K$ [ms$^{-1}$] & $181.0_{-5.4}^{+5.5}$ $\dagger$ & - \\
$\gamma$ [ms$^{-1}$] & $-29.2_{-2.6}^{+2.6}$ $\dagger$ & - \\
$B$ & $1.04246 \pm 0.00023$ $\dagger$ & $1^{*}$ \\
$u_1$ & $0.52_{-0.39}^{+0.18}$ & $0.22^{*}$ \\
$u_2$ & $0^{\mathrm{*}}$ & $0.32^{*}$ \\
$R_P/R_*$ & $0.1294_{-0.0046}^{+0.0020}$ & $0.1253 \pm 0.0010$ \\
$a/R_*$ & $7.93_{-0.24}^{+0.18}$ & $7.63 \pm 0.12$ \\
$i$ [$^{\circ}$] & $83.99_{-0.35}^{+0.25}$ & $83.57 \pm 0.14$ \\
$\rho_*$ [g\,cm$^{-3}$] & $1543.4_{-135.6}^{+108.7}$ & - \\
$\log(g_P\,[\mathrm{cgs}])$ & $3.306_{-0.018}^{+0.018}$ & - \\ [1ex]
\hline\hline 
\end{tabular}
\label{tab:holman} 
\end{table}

\section{SUMMARY OF RESULTS}

Due to the large number of results presented in this paper, we summarize the key 
findings below:

\begin{itemize}
\item[{\tiny$\blacksquare$}] The \emph{Kepler} SC data exhibit unprecedented 
precision with r.m.s. noise 237.2\,ppm per 58.8\,s.
\item[{\tiny$\blacksquare$}] Fitting for limb darkening coefficients leads to 
much larger uncertainties in the system parameters of TrES-2b, due to the 
near-grazing nature of the orbit (e.g. a factor of 17.5 larger for the transit 
depth)
\item[{\tiny$\blacksquare$}] We present a self-consistent, refined set of 
transit, radial velocity and physical parameters for the TrES-2b system, which 
are in close agreement with previous values.
\item[{\tiny$\blacksquare$}] We do not detect an occultation of TrES-2b,
constraining the depth to be $<72.9$\,ppm to 3-$\sigma$ confidence, indicating 
that this object has the lowest measured geometric albedo for an exoplanet, of 
$A_g < 0.146$.
\item[{\tiny$\blacksquare$}] We detect no short or long term transit timing 
variations (TTV) in the TrES-2b system and exclude short-term signals of r.m.s. 
$>7.11$\,s and a long-term variation of $0.11$\,s per year in the orbital 
period, to 3-$\sigma$ confidence.
\item[{\tiny$\blacksquare$}] We detect no short or long term transit duration
variations (TDV) in the TrES-2b system and exclude short-term relative duration 
change of $>0.77$\% and long-term change of $>2.5$\% per year, to 3-$\sigma$ 
confidence.
\item[{\tiny$\blacksquare$}] We exclude the presence of exomoons down to 
sub-Earth masses for TrES-2b.
\item[{\tiny$\blacksquare$}] The \citet{mislis:2009} hypothesis of long-term
duration change is neither supported nor refuted by our analysis.
\item[{\tiny$\blacksquare$}] We find the \citet{rabus:2009} hypothesis of a 
0.2\,cycle TTV is not supported by the \emph{Kepler} photometry, to a high
confidence level.
\item[{\tiny$\blacksquare$}] We find no evidence for other dips in the 
lightcurve as reported by \citet{raetz:2009}.
\end{itemize}

\acknowledgements
\section*{Acknowledgements}

We would like to thank the \emph{Kepler} Science Team and everyone who
contributed to making the \emph{Kepler Mission} possible. We are
extremely grateful to the \emph{Kepler} Science Team and the Data Analysis 
Working Group for making the reduced photometry from Q0 and Q1 publicly 
available. Thanks to J. Jenkins, R. Gilliland, J. Winn, D. Latham, 
J. P. Beaulieu, G. Tinetti, J. Steffen and J. Rowe for useful comments in 
preparing this manuscript. We are also very grateful to the anonymous referee
for their helpful advise and feedback.

D.M.K.~has been supported by STFC studentships, Smithsonian Institution FY11
Sprague Endowment Funds and by the HATNet as an SAO predoctoral fellow. We 
acknowledge NASA NNG04GN74G, NNX08AF23G grants, and Postdoctoral Fellowship of 
the NSF Astronomy and Astrophysics Program (AST-0702843 for G.~B.). Special 
thanks to the amateur astronomy community and the ETD.




\clearpage

\begin{table*}
\caption{\emph{All measured transit times of TrES-2b taken from the literature, 
the ETD (Exoplanet Trasnit Database) and this work, at the time of writing. 
* = value presented in \citet{mislis:2009} and \citet{mislis:2010} for this 
transit do not agree with each other and therefore these measurements are not 
included in our long-term TTV analysis. $\dagger$ = value comes from summing 
more than one transit lightcurve. ** = times are in BJD$_{\mathrm{TDB}}$.}} 
\centering 
\begin{tabular}{c c c c c c} 
\hline\hline 
Epoch & $\tau$ [HJD$_{\mathrm{UTC}}$-2,450,000] & Reference & Epoch & $\tau$ [HJD$_{\mathrm{UTC}}$-2,450,000] & Reference \\ [0.5ex] 
\hline
000 & $3957.63580 \pm 0.00100$ & \citet{donovan:2007} & 391 & $4923.64380 \pm 0.00070$ & ETD \\
004 & $3967.51800 \pm 0.00043$ & \citet{rabus:2009} & 393 & $4928.58752 \pm 0.00143$ & ETD \\
012 & $3987.28000 \pm 0.00800$ & ETD & 393 & $4928.58757 \pm 0.00626$ & ETD \\
013 & $3989.75286 \pm 0.00029$ & \citet{holman:2007} & 393 & $4928.58792 \pm 0.00112$ & ETD \\
015 & $3994.69393 \pm 0.00031$ & \citet{holman:2007} & 395 & $4933.52740 \pm 0.00076$ & \citet{mislis:2010} \\
019 & $4004.57500 \pm 0.00140$ & ETD & 395 & $4933.52726 \pm 0.00168$ & ETD \\
025 & $4019.40150 \pm 0.00600$ & ETD & 399 & $4943.41320 \pm 0.00138$ & ETD \\
034 & $4041.63579 \pm 0.00030$ & \citet{holman:2007} & 404 & $4955.763285_{-0.000055}^{+0.000055}$ & This work ** \\
087 & $4172.57670 \pm 0.00160$ & \citet{raetz:2009} & 405 & $4958.233958_{-0.000056}^{+0.000056}$ & This work ** \\
106 & $4219.52050 \pm 0.00600$ & ETD & 406 & $4960.704556_{-0.000054}^{+0.000054}$ & This work ** \\
108 & $4224.46176 \pm 0.00250$ & \citet{raetz:2009} & 407 & $4963.175188_{-0.000056}^{+0.000056}$ & This work ** \\
127 & $4271.39911 \pm 0.00297$ & ETD & 408 & $4965.645708_{-0.000054}^{+0.000054}$ & This work ** \\
130 & $4278.81790 \pm 0.00600$ & ETD & 409 & $4968.116367_{-0.000053}^{+0.000053}$ & This work ** \\
138 & $4298.57880 \pm 0.00240$ & \citet{raetz:2009} & 410 & $4970.587001_{-0.000056}^{+0.000056}$ & This work ** \\
140 & $4303.52090 \pm 0.00030$ & \citet{rabus:2009} & 410 & $4970.58650 \pm 0.00100$ & ETD \\
142 & $4308.46130 \pm 0.00045$ & \citet{rabus:2009} & 411 & $4973.057600_{-0.000057}^{+0.000056}$ & This work ** \\
142 & $4308.46448 \pm 0.00130$ & \citet{raetz:2009} & 412 & $4975.528262_{-0.000058}^{+0.000059}$ & This work ** \\
142 & $4308.46240 \pm 0.00600$ & ETD & 412 & $4975.52630 \pm 0.00150$ & ETD \\
142 & $4308.46300 \pm 0.00180$ & ETD & 412 & $4975.52790 \pm 0.00090$ & ETD \\
151 & $4330.70130 \pm 0.00200$ & ETD & 413 & $4977.998830_{-0.000057}^{+0.000057}$ & This work ** \\
155 & $4340.58350 \pm 0.00120$ & ETD & 414 & $4980.469389_{-0.000056}^{+0.000056}$ & This work ** \\
157 & $4345.51390 \pm 0.00160$ & ETD & 414 & $4980.46750 \pm 0.00060$ & \citet{mislis:2010} $\dagger$ \\
157 & $4345.51990 \pm 0.00120$ & ETD & 414 & $4980.46450 \pm 0.00130$ & ETD \\
157 & $4345.52350 \pm 0.00150$ & ETD & 414 & $4980.46790 \pm 0.00170$ & ETD \\
163 & $4360.34550 \pm 0.00109$ & \citet{raetz:2009} $\dagger$ & 414 & $4980.46820 \pm 0.00130$ & ETD \\
165 & $4365.28746 \pm 0.00210$ & \citet{raetz:2009} & 415 & $4982.939972_{-0.000058}^{+0.000058}$ & This work ** \\
170 & $4377.63810 \pm 0.00070$ & ETD & 416 & $4985.410672_{-0.000055}^{+0.000055}$ & This work ** \\
170 & $4377.64230 \pm 0.00120$ & ETD & 417 & $4987.881247_{-0.000055}^{+0.000055}$ & This work ** \\
174 & $4387.52220 \pm 0.00150$ & \citet{raetz:2009} & 418 & $4990.351850_{-0.000056}^{+0.000056}$ & This work ** \\
229 & $4523.40970 \pm 0.00080$ & ETD & 419 & $4992.822571_{-0.000055}^{+0.000056}$ & This work ** \\
242 & $4555.52621 \pm 0.00123$ & ETD & 420 & $4995.293071_{-0.000057}^{+0.000056}$ & This work ** \\
242 & $4555.52360 \pm 0.00090$ & ETD & 421 & $4997.763671_{-0.000058}^{+0.000059}$ & This work **\\
259 & $4597.52250 \pm 0.00120$ & ETD & 421 & $4997.76286 \pm 0.00035$ & ETD \\
263 & $4607.40360 \pm 0.00720$ & \citet{mislis:2009} & 423 & $5002.70200 \pm 0.00090$ & ETD \\
268 & $4619.75990 \pm 0.00130$ & ETD & 425 & $5007.64270 \pm 0.00190$ & ETD \\
272 & $4629.64510 \pm 0.00240$ & ETD & 429 & $5017.52520 \pm 0.00100$ & ETD \\
274 & $4634.58280 \pm 0.00030$ & \citet{rabus:2009} & 433 & $5027.40740 \pm 0.00190$ & ETD \\
276 & $4639.52320 \pm 0.00031$ & \citet{rabus:2009} & 438 & $5039.76060 \pm 0.00110$ & ETD \\
278 & $4644.46608 \pm 0.00140$ & \citet{raetz:2009} & 438 & $5039.76480 \pm 0.00070$ & ETD \\
278 & $4644.46440 \pm 0.00180$ & ETD & 438 & $5039.76607 \pm 0.00096$ & ETD \\
280 & $4649.41490 \pm 0.00330$ & ETD & 438 & $5039.76680 \pm 0.00120$ & ETD \\
281 & $4651.87560 \pm 0.00070$ & ETD & 440 & $5044.70310 \pm 0.00080$ & ETD \\
293 & $4681.52240 \pm 0.00210$ & ETD & 442 & $5049.64530 \pm 0.00120$ & ETD \\
304 & $4708.69870 \pm 0.00110$ & ETD & 442 & $5049.64940 \pm 0.00085$ & ETD \\
310 & $4723.51790 \pm 0.00190$ & ETD & 446 & $5059.52244 \pm 0.00076$ & ETD \\
312 & $4728.47400 \pm 0.00710$ & \citet{mislis:2009} * & 548 & $5311.53095 \pm 0.00077$ & ETD \\
312 & $4728.46400 \pm 0.00710$ & \citet{mislis:2010} * & 550 & $5316.47653 \pm 0.00094$ & ETD \\
316 & $4738.35215 \pm 0.00200$ & \citet{raetz:2009} & 552 & $5321.41833 \pm 0.00137$ & ETD \\
316 & $4738.35045 \pm 0.00090$ & ETD & 555 & $5328.82558 \pm 0.00093$ & ETD \\
318 & $4743.28972 \pm 0.00180$ & \citet{raetz:2009} & 557 & $5333.76390 \pm 0.00107$ & ETD \\
321 & $4750.70010 \pm 0.00110$ & ETD & 557 & $5333.76469 \pm 0.00123$ & ETD \\
333 & $4780.34690 \pm 0.00220$ & ETD & 567 & $5358.47237 \pm 0.00071$ & ETD \\ [1ex]
\hline\hline 
\end{tabular}
\label{tab:allttv} 
\end{table*}

\end{document}